\documentclass[twocolumn,prx,twocolumn,superscriptaddress]{revtex4}
\usepackage[latin9]{inputenc}
\usepackage{color}
\usepackage{amsmath}
\usepackage{amssymb}
\usepackage{graphicx}
\usepackage{tikz}
\usepackage{verbatim}
\usepackage{amsmath}
\usepackage{amssymb}
\usepackage{graphicx}
\usepackage{tikz}
\usepackage{amsmath,bm,amssymb,latexsym,galois,amsthm,euscript}
\usepackage[all,cmtip,knot]{xy}
\usepackage{mathrsfs}
\usepackage[unicode=true,bookmarks=true,bookmarksnumbered=false,bookmarksopen=false,breaklinks=false,pdfborder={0 0 1},backref=false,colorlinks=true]{hyperref}

\usepackage{overpic}
\usepackage{algpseudocode}
\usepackage{algorithm}

\hypersetup{
    colorlinks,
    linkcolor={red!50!black},
    citecolor={blue!90!black},
    urlcolor={blue!90!black}
}

\makeatletter
\@ifundefined{textcolor}{}
{%
 \definecolor{BLACK}{gray}{0}
 \definecolor{WHITE}{gray}{1}
 \definecolor{RED}{rgb}{1,0,0}
 \definecolor{GREEN}{rgb}{0,1,0}
 \definecolor{BLUE}{rgb}{0,0,1}
 \definecolor{CYAN}{cmyk}{1,0,0,0}
 \definecolor{MAGENTA}{cmyk}{0,1,0,0}
 \definecolor{YELLOW}{cmyk}{0,0,1,0}
}

\usepackage{amsfonts}\usepackage{tabularx}\usepackage{dcolumn}\usepackage{bm}\usepackage{graphicx}\usepackage{epstopdf}

\hypersetup{urlcolor=blue}

\newcommand{\Cb}{\mathbb{C}}
\newcommand{\Zb}{\mathbb{Z}}

\theoremstyle{definition}

\newtheorem*{defn*}{Definition}

\theoremstyle{remark}

\makeatother

\begin{document}

\title{Efficient Machine Learning Representations of Surface Code with Boundaries, Defects, Domain Walls and Twists}
\author{Zhih-Ahn Jia\footnote{Two authors are of equal contributions.}}
\email{\tt giannjia@foxmail.com}
\affiliation{Microsoft Station Q, University of California, Santa Barbara, California 93106-6105, USA}
\affiliation{Department of Mathematics, University of California, Santa Barbara, California 93106-6105, USA}
\affiliation{Key Laboratory of Quantum Information, Chinese Academy of Sciences, School of Physics, University of Science and Technology of China, Hefei, Anhui, 230026, P.R. China}
\affiliation{Synergetic Innovation Center of Quantum Information and Quantum Physics, University of Science and Technology of China, Hefei, Anhui, 230026, P.R. China}
\author{Yuan-Hang Zhang\footnotesize{*}}
\affiliation{School of the Gifted Young, University of Science and Technology of China, Hefei, Anhui, 230026, P.R. China}
\author{\normalsize{Yu-Chun Wu}}
\email{\tt wuyuchun@ustc.edu.cn}
\affiliation{Key Laboratory of Quantum Information, Chinese Academy of Sciences, School of Physics, University of Science and Technology of China, Hefei, Anhui, 230026, P.R. China}
\affiliation{Synergetic Innovation Center of Quantum Information and Quantum Physics, University of Science and Technology of China, Hefei, Anhui, 230026, P.R. China}

\author{Liang Kong}
\email{\tt kongl@sustc.edu.cn}
\affiliation{Shenzhen Institute for Quantum Science and Engineering, and Department of Physics, Southern University of Science and Technology, Shenzhen 518055, China}

\author{Guang-Can Guo}
\affiliation{Key Laboratory of Quantum Information, Chinese Academy of Sciences, School of Physics, University of Science and Technology of China, Hefei, Anhui, 230026, P.R. China}
\affiliation{Synergetic Innovation Center of Quantum Information and Quantum Physics, University of Science and Technology of China, Hefei, Anhui, 230026, P.R. China}
\author{Guo-Ping Guo}
\affiliation{Key Laboratory of Quantum Information, Chinese Academy of Sciences, School of Physics, University of Science and Technology of China, Hefei, Anhui, 230026, P.R. China}
\affiliation{Synergetic Innovation Center of Quantum Information and Quantum Physics, University of Science and Technology of China, Hefei, Anhui, 230026, P.R. China}
\begin{abstract}
Machine learning representations of many-body quantum states have recently been introduced as an ansatz to describe the ground states and unitary evolutions of many-body quantum systems. We investigate one of the most important representations, restricted Boltzmann machine (RBM), in stabilizer formalism. A general method to construct RBM representations for
stabilizer code states is given and exact RBM representations for several types of stabilizer groups
with the number of hidden neurons equal or less than the number of visible neurons are presented. The result indicates
that the representation is extremely efficient. Then we analyze the surface code with boundaries,
defects, domain walls and twists in full detail and find that almost all the models can be efficiently represented via RBM ansatz: the RBM parameters of perfect case, boundary case, and defect case are constructed analytically using the method we provide in stabilizer formalism; and the domain wall and twist case is studied numerically.  Besides, the case for Kitaev's $D(\Zb_d)$ model, which is a generalized model of surface code, is also investigated.
\end{abstract}
\maketitle

\section{Introduction}
\label{sec:RBM}
As the leading proposal of achieving fault-tolerant quantum computation, surface code attracts researchers' great attentions in recent years. Since Kitaev \cite{Kitaev2003} made an ingenious step of transforming a quantum error correction code (QECC) into a many-body interacting quantum system (more precisely, he constructed a Hamiltonian, now known as toric code, which is a gapped anyon system and whose ground state space is exactly the code space, the encoded information is protected by the topological properties of the system), surface code model has been extensively investigated from both QECC perspective and condensed matter perspective. The studies on surface code cross-fertilize both areas. Suppose that we are encoding information with $n$ physical bits, i.e., with the Hilbert space $\mathcal{H}=(\mathbb{C}^2)^{\otimes n}$. The code space $\mathcal{C}$ is a subspace of $\mathcal{H}$. This subspace can be regarded as the ground state space of the corresponding surface code Hamiltonian $H$, which is just the negative of the summation of all stabilizer generators.

From condensed matter perspective, to conquer the challenge of efficiently representing the many-body states with exponential complexity, neural network, one of the most important tools in machine learning \cite{lecun2015deep,Hinton2006}, is introduced to efficiently represent the ground states of strongly correlated many-body systems \cite{Carleo602}, which are beyond the mean-field paradigm, for which the density matrices are of the tensor product form in the thermodynamic limit. Mean-field approach is very successful in bosonic system (quantum de Finetti theorem) but fails for other strongly correlated system. There are also some other approaches: quantum Monte Carlo method \cite{Scalapino1981,Blankenbecler1981,fucito1980proposal,Hirsch1985,Hirsch1981}, which suffers from the sign problem, and tensor network representation \cite{ORUS2014}, whose special form, matrix product states (MPS), makes a great success in $1 d$ system \cite{landau2015polynomial,Arad2017},  but for $2 d$ case, it is unknown whether the corresponding projected entangled pair states (PEPS)  are enough and extracting information is $\#P$-hard in general, the best known approximation algorithm still spends superpolynomial time under assumptions \cite{Schuch2007,Anshu2016}. The connections between machine learning representation and other representations are also extensively exploited \cite{Chen2018,Huang2017a,gao2017efficient,Glasser2018}.

Machine learning as a method for analyzing data has been prevalent in many scientific areas \cite{lecun2015deep,Hinton2006,sutton1998reinforcement}, including computer vision, speech recognition, chemical synthesis, etc. Among which artificial neural network plays an important role in recognizing or even discovering particular patterns of the input data. Quantum machine learning (QML) \cite{biamonte2017quantum}, which is an emerging interdisciplinary scientific area at the intersection of quantum mechanics and machine learning, has recently attracted many attentions \cite{Rebentrost2014,Dunjko2016,Monras2017,carrasquilla2017machine,Carleo602,Deng2017,Deng2017a,gao2017efficient}. There are two crucial branches of QML, the first one is to develop new quantum algorithms which share some features of machine learning and behave faster and better than their classical counterparts \cite{Rebentrost2014,Dunjko2016,Monras2017}; the second one, which is also the focus of this work, is to use the classical machine learning methods to assist the study of quantum systems. Machine learning methods are so powerful, that it can be used for distinguishing phases \cite{carrasquilla2017machine}, quantum control \cite{August2017}, error-correcting of topological codes \cite{Torlai2017}, quantum tomography \cite{Zhang2017,torlai2017many} and efficiently representing quantum many-body states \cite{Carleo602,Deng2017,Deng2017a,gao2017efficient,huang2017neural,saito2017solving,Cai2018}. Among all of them, using neural network as the variational wave functions to approximate ground state of many-body quantum systems received many attentions recently. Many different neural network architectures are tested and the most successful one is restricted Boltzmann machine (RBM) \cite{Carleo602,Deng2017,Deng2017a,gao2017efficient}. It has been shown that RBM can efficiently represent ground states of several many-body models, including Ising model \cite{Carleo602}, toric code model \cite{Deng2017,Deng2017a} and graph states \cite{gao2017efficient}.

In this work, we study the RBM representation in stabilizer formalism and we provide some more systematic analyses. It is shown that for many stabilizer groups, the RBM representations are extremely efficient: the number of hidden neurons approximates the number of visible neurons. We take the surface code with boundaries, defects, domain walls and twists as some concrete examples, and we find all these models can be represented by RBM. The exact solution is given for the boundary and defect cases.  We also analyze the Kitaev's $D(G)$ model for $G=\Zb_d$ case. All these models are investigated for the first time, and our results can be useful for building the RBM neural network when analyzing anyon model or QECC in stabilizer formalism.

The work is organized as follows. In Sec. \ref{sec:surfaceCODE} we provide an elaborate description of stabilizer formalism, Kitaev's $D(G)$ model, $\Zb_2$-surface code model, and the properties when these models are regarded as anyon models. Then we construct the surface code models with boundaries, defects, domain walls and twists, and give the precise stabilizer operators and Hamiltonians of these models. In Sec. \ref{sec:RBMreview}, we give a brief review of RBM representations of states. Then, in Sec. \ref{sec:ML-rep}, the RBM representations in stabilizer formalism are worked out and many explicit solutions of stabilizer states are constructed. In Sec. \ref{sec:RMB-rep}, using the results developed in Sec. \ref{sec:ML-rep}, we provide a detailed analysis of RBM representations of surface code with  boundaries, defects, domain walls and twists. And the general Kitaev's $D(G)$ case is done in Sec. \ref{sec:DG}. Finally, in Sec. \ref{sec:conclusion} we make some discussions and give the conclusions.

\section{Surface code model with boundaries, defects, domain walls and twists}
\label{sec:surfaceCODE}

In this section, we give a brief review of the basics of surface code in stabilizer formalism, and the corresponding surface code Hamiltonian is an anyon model. For simplicity of illustration, we will assume hereinafter that the lattice is square lattice placed on plane, but all our results can be extended to general cases similarly. We will analyze the boundaries, defects, domain walls and twists in surface code from anyon theoretic perspective.

\subsection{Stabilizer formilsm}

QECCs are commonly expressed in the stabilizer formalism \cite{gottesman1997stabilizer,Gottesman1998}. To prevent the encoded information from noise, the logical quantum states are encoded redundantly in a $k$-dimensional subspace $\mathcal{C}$ of the $n$-qubit physical space $\mathcal{H}=(\mathbb{C}^2)^{\otimes n}$. The stabilizer group $\mathbf{S}$  for $\mathcal{C}$ is an Abelian subgroup of Pauli group $\mathbf{P}_n=\{I,\sigma_x,\sigma_y,\sigma_z\}^{\otimes n}\times \{\pm 1,\pm i\}$, more precisely, $\mathcal{C}$ is the invariant subspace for $\mathbf{S}$ acting on $\mathcal{H}$. Since each operator $T_j$ in $\mathbf{S}$ is a Hermitian operator and $[T_i,T_j]=0$ for all $i,j$, the code states are the common eigenstates of all elements $T_j$ in $\mathbf{S}$, i.e.,
\begin{equation}\label{eq:stabilizer}
T_j|\Psi\rangle =+1 |\Psi\rangle,\,\, \forall j.
\end{equation}

Suppose $\mathbf{S}$ is generated by $m$ independent operators $\{T_1,\cdots,T_m\}$. Note that $T_j^2=I$ for all $j=1,\cdots,m$, then  any $T\in\mathbf{S}$ can be uniquely expressed as $T=T_1^{\alpha_1}T_{2}^{\alpha_2}\cdots T_{m}^{\alpha_m}$ where $\alpha_j\in\{0,1\}$, thus the order of stabilizer group $\mathbf{S}$ is $2^m$. The numbers of physical qubits $n$, generators of stabilizer group $m$ and encoded logical qubits $k$ are related by a simple formula $n=m+k$.

To construct logical operators $\bar{L}$ which leave the code space invariant and transform the logical states into each other, notice that any pair of Pauli operators must commute or anticommute. Any Pauli operator anticommute with elements in $\mathbf{S}$ can not leave the code space invariant and logical gates must not be in $\mathbf{S}$ or they can not achieve the logical transformation. Therefore, logical gates operator must live in the centralizer $\mathbf{C}\subset \mathbf{P}_n$ of the stabilizer group. It's worth mentioning that the representation of logical operator is not unique. Two logical operators $\bar{L}$ and $\bar{L}'=\bar{L}T$ with $T\in\mathbf{S}$ satisfy $\bar{L}'|\Psi\rangle = \bar{L}|\Psi\rangle$ for all code states $|\Psi\rangle$.

Another important quantity to characterize stabilizer code is the code distance $d$. It is defined as the smallest set of qubits which supports one nontrivial logical operator of the code. The stabilizer code with $n$ physical qubits, $k$ encoded logical qubits and code distance $d$ is denoted as $[[n,k,d]]$.

\begin{figure}
\includegraphics[scale=0.28]{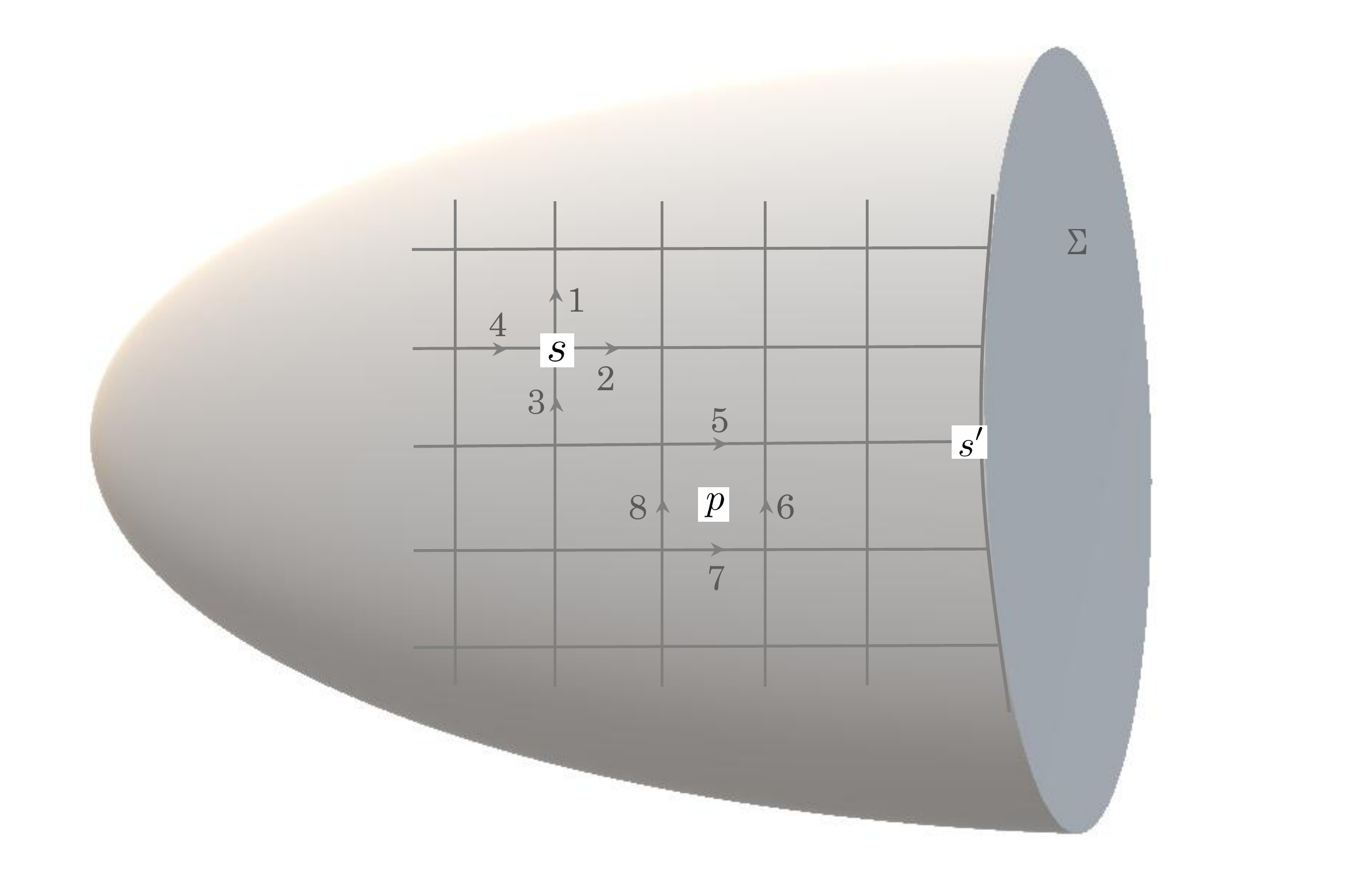}
\caption{\label{fig:surface}Surface code model on a surface $\Sigma$.}
\end{figure}

\subsection{Lattice model on a surface}

The anyon model of surface code is a $D(\mathbb{Z}_2)$ quantum double model \cite{Kitaev2003}. For a given surface $\Sigma$, consider its cellulation $\mathcal{C}(\Sigma)$ which is the set of all cells, we denote the set of 2-cells (i.e., plaquettes) as $\mathcal{C}^2(\Sigma)$, 1-cells (i.e., edges) $\mathcal{C}^1(\Sigma)$ and 0-cells (i.e. vertices) $\mathcal{C}^0(\Sigma)$. We can attach a physical space $\mathcal{H}_{e_i}$ on each edge $e_i$ of the lattice, the basis is chosen as $\{|g\rangle: g\in G\}$ labeled by elements in $G$, the whole space is then $\mathcal{H}=\bigotimes_{e_i\in \mathcal{C}^1(\Sigma)}\mathcal{H}_{e_i}$. Quantum double model $D(G)$ for general finite group $G$ can be defined on general two-dimensional lattice, but here, for convenience we only employ the square lattice and the group $G$ is chosen as Abelian group $\Zb_2$. To proceed we define the operators $L^{g}_{\pm}$ which are associated with the vertices of the lattice $\mathcal{C}^1(\Sigma)$ and $T_{\pm}^h$ which are associated with plaquettes of the lattice $\mathcal{C}^2(\Sigma)$, such that
\begin{align}\label{eq:LT}
L^{g}_{+}|z\rangle= |gz\rangle,\,\,\, & \,\,\, L^{g}_{-}|z\rangle= |zg^{-1}\rangle,\nonumber \\
T^{h}_{+}|z\rangle= \delta_{h,z}|z\rangle, \,\,\,& \,\,\, T^{h}_{-}|z\rangle= \delta_{h^{-1},z}|z\rangle.\nonumber
\end{align}
It is easy to check that these operators satisfy the following relations:
\begin{equation}\label{}
  \begin{split}\label{}
   L_{+}^{g} T_{+}^{h}=T_{+}^{gh}L_{+}^{g}, \,\,\, & \,\,\,  L_{+}^{g} T_{-}^{h}=T_{-}^{hg^{-1}}L_{+}^{g}, \nonumber\\
   L_{-}^{g} T_{+}^{h}=T_{+}^{hg^{-1}}L_{+}^{g}, \,\,\, & \,\,\,  L_{-}^{g} T_{-}^{h}=T_{-}^{gh}L_{-}^{g}. \nonumber
  \end{split}
\end{equation}

Now consider the orientable surface $\Sigma$ as in Fig. \ref{fig:surface}, where a square lattice is placed on it. To consistently define the Hamiltonian we give each edge an orientation, here we take the vertical edges upwards and horizontal edges rightwards. Changing of the orientation of edge corresponds to changing $|g\rangle$ to $|g^{-1}\rangle$. For $\Zb_2$ case, $0^{-1}=0$ and $1^{-1}=1$, thus the orientation is not necessary for $D(\Zb_2)$ model. We now define two types of operators, star operators defined on vertices (see $s$ vertex as in Fig. \ref{fig:surface})
\begin{equation}\label{eq:vertex}
  A(s)=\frac{1}{|G|}\sum_{g\in G}L_{-,1}^{g}L_{-,2}^{g}L_{+,3}^{g}L_{+,4}^{g},
\end{equation}
where for edges pointing to $s$ we assign $L_{+}^{g}$, otherwise we assign $L_{-}^{g}$. And plaquette operators is defined as (see $p$ plaquette as in Fig. \ref{fig:surface})
\begin{equation}\label{eq:plaquette}
  B(p)=\sum_{h_5h_6h_7h_8=1_G}T^{h_5}_{-,5}T^{h_6}_{+,6}T^{h_7}_{+,7}T^{h_8}_{-,8},
\end{equation}
where if $p$ is on the left of edge we assign $T^{h}_{+}$ to the edge, otherwise we assign $T^{h}_{-}$. $A(s)$, $A(s')$, $B(p)$, $B(p')$ commute with each other for all $s,s'\in\mathcal{C}^{0}(\Sigma)$ and $p,p'\in\mathcal{C}^{2}(\Sigma)$.

Note that for group $\Zb_2$, $0^{-1}=0$ and $1^{-1}=1$, thus $L_{+}^0=L_{-}^0=I$ and $L_{+}^1=L_{-}^1=\sigma_x$; and $T_{+}^0=T_{-}^0=\Pi_{|0\rangle\langle 0|}$ and $L_{+}^1=L_{-}^1=\Pi_{|1\rangle\langle 1|}$. We introduce new operators $A_s$ and $B_p$. For each vertex (star) $s$ and each plaquette $p$, construct the following vertex and plaquette operators
\begin{equation}\label{}
 A_s = \Pi_{i \in star(s)} \sigma^i_x~~~~~~~B_p = \Pi_{i \in \partial p} \sigma^i_z,
\end{equation}
where we use $\partial p$ to represent the boundary edges of the plaquette $p$. $A_s$ and $B_p$ are stabilizer operators in stabilizer QECC formalism and they commute with each other, i.e., $[A_s,A_{s'}]=[B_p,B_{p'}]=[A_s,B_p]=0$ for all vertices $s,s'$ and plaquette $p,p'$. And all these operators are Hermitian with eigenvalues $\pm 1$. Notice that $A(s)=\frac{1}{2}(I+A_s)$ and $B(p)=\frac{1}{2}(I+B_p)$. Here for simplicity we construct the following Hamiltonian
\begin{equation}\label{}
  H_{\Sigma}=-\sum_{s}A_s -\sum_p B_p,
\end{equation}
which will be referred to as the surface code Hamiltonian. $H_{\Sigma}$ is the negative summation of all stabilizer generators, thus the ground states for it corresponds to the solution $A_{s}|\Omega\rangle=|\Omega\rangle$ $B_{p}|\Omega\rangle=|\Omega\rangle$ for all $s\in \mathcal{C}^{0}(\Sigma)$ and $p\in \mathcal{C}^{2}(\Sigma)$, which turn out to be code states in stabilizer formalism.

\subsection{Boundaries, defects, and twists}
\begin{figure}
\includegraphics[width=8.8cm]{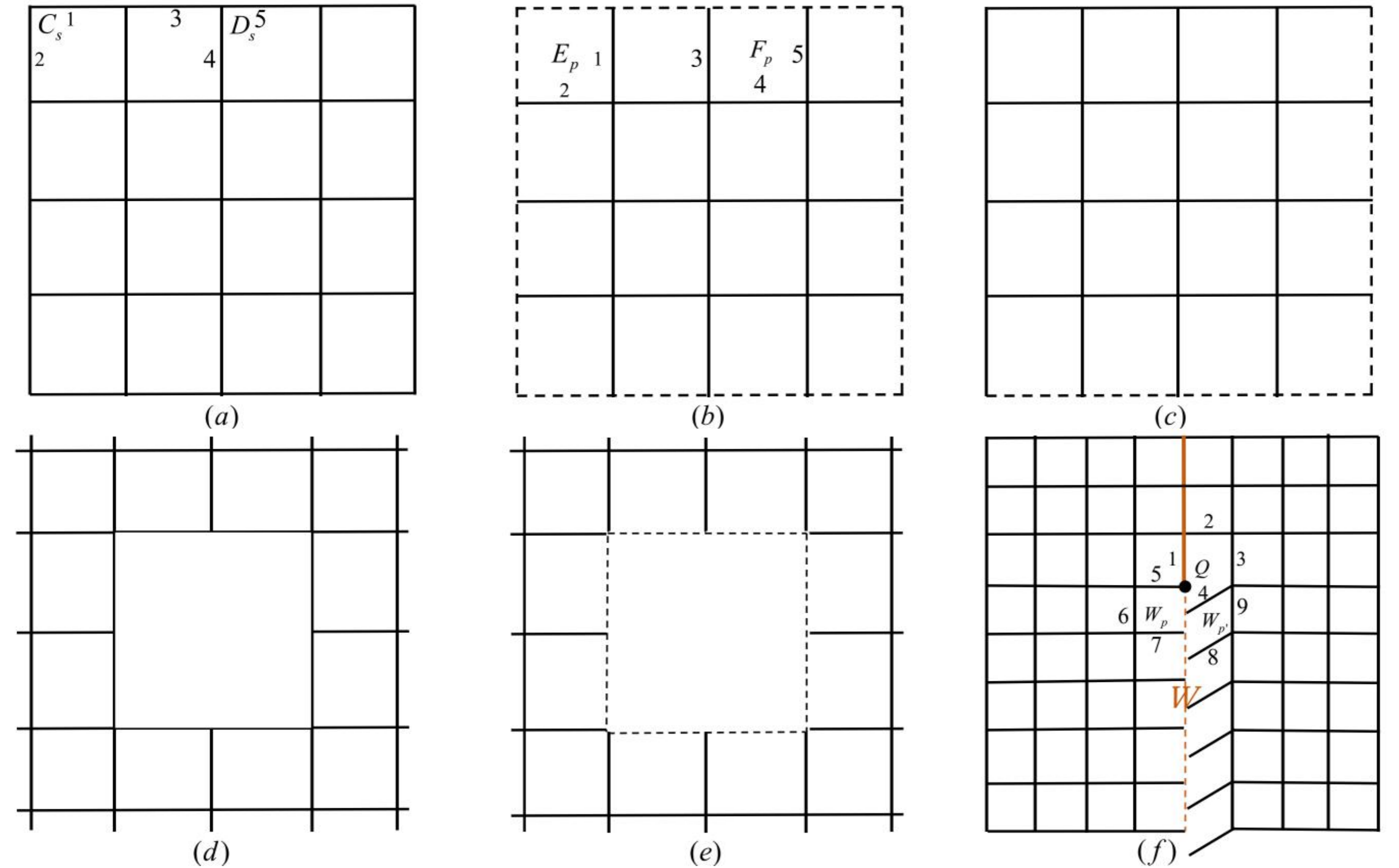}
\caption{\label{fig:boundary}The surface code with boundaries, defects and twists. (a) smooth boundary; (b) rough boundary; (c) mixed boundary; (d) smooth defect; (e) rough defect; (f) twist $Q$ and domain wall $W$.}
\end{figure}

Real samples of quantum matter have boundaries and defects, so it is also important to analyze the quantum double model on a lattice with boundaries and defects. As discussed in Refs. \cite{Kitaev2012a,Beigi2011,Bombin2008,Cong2017}, we can construct boundary and defect Hamiltonians. For convenience, we denote $H_{bulk}$, $H_{bondary}$, $H_{defect}$ and $H_{twist}$ the Hamiltonians of bulk, boundaries, defects and twists respectively.
\subsubsection{Gapped boundaries of surface code}
In general, the gapped boundary of the quantum double model $D(G)$ is determined by the subgroup $K \in G$ (up to conjugation) and a $2$-cocycle in $H^{2}(K,\Cb^{\times})$ \cite{Beigi2011}. To define a Hamiltonian for gapped boundaries, we first need to give the orientation of the boundaries and then introduce the local terms of each star and plaquette near the boundary (which depend on a subgroup $K$ of $G$). Here, we will focus on the simplest $\Zb_2$ case and we take $K$ to be $\Zb_2$ itself.

There are two types of boundaries for planar code: smooth one and rough one \cite{bravyi1998quantum,freedman2001projective}. Let us now define some new star and plaquette operators (see Fig. \ref{fig:boundary} (a)-(c)). For smooth boundaries, we can see that the plaquettes does not change near the boundary, but the star operators changes. We need to introduce two kinds of operators: corner star operator $C_s$ and boundary star operator $D_s$ (see Fig. \ref{fig:boundary} (a)):
\begin{equation*}\label{}
C_s=\sigma_x^1\sigma_x^2,\,\,\, D_s=\sigma_x^3\sigma_x^4\sigma_x^5.
\end{equation*}
The smooth boundary Hamiltonians then reads:
\begin{align}\label{}
H_{sboundary}=-\sum_{s}C_s -\sum_s D_s,\nonumber
\end{align}
in which all terms are commutative with each other, thus $H_{sboundary}$ is gapped Hamiltonian.

For the rough boundaries, the star operator near the boundary remain unchanged but the plaquette operators change, and we similarly introduce two kinds of operators: corner  plaquette operator $E_p$ and boundary plaquette operator $F_p$ (see Fig. \ref{fig:boundary} (b)):
\begin{equation*}\label{}
E_p=\sigma_z^1\sigma_z^2,\,\,\, F_p=\sigma_z^3\sigma_z^4\sigma_z^5.
\end{equation*}
Similarly, we have the gapped Hamiltonian for rough boundaries
\begin{align}\label{}
H_{rboundary}=-\sum_{p}E_p -\sum_p F_p.\nonumber
\end{align}

We can also introduce the mixed boundaries which is the mixed case of smooth and rough boundaries (see Fig. \ref{fig:boundary} (c)). The boundary Hamiltonian then reads:
\begin{align}\label{}
H_{mboundary}=&-\sum_{s}C_s -\sum_s D_s,\nonumber\\
 &-\sum_{p}E_p -\sum_p F_p.\nonumber
\end{align}

When an $e$ particle move to the rough  boundary, it will condense into the vacuum of the boundary. Similarly, $m$ particle will condense in the smooth boundary. Thus the boundary phase is condensed from the bulk phase. Conversely, the bulk phase can also be recovered from the boundary phase via the half-loop of the $m$ and $e$ particles. This is the content of the famous boundary-bulk duality.

\subsubsection{Defects of surface code}
Let us now consider the case where we punch several holes $\mathfrak{h}_1,\cdots,\mathfrak{h}_k$ on the lattice. To describe the holes, we need to specify $k$ subgroups $K_1,\cdots, K_k$ of $G$. Here, we still assume that all $K_i$ are equal to $G=\Zb_2$. Then the hole Hamiltonian will be
\begin{equation*}\label{}
H_{defect}=\sum_i H_{\mathfrak{h}_i}.
\end{equation*}
Like the case for boundaries, there are two typical types of holes: smooth one and rough one, see Fig. \ref{fig:boundary} (d) and (e). The main difference is that we do not need to introduce the corner star operator for smooth hole, and do not need to introduce corner plaquette operators for rough hole. Therefore, we have the Hamiltonians for holes as:
\begin{align*}
  H_{s\mathfrak{h}}=-\sum_s D_s,\,\,\, H_{r\mathfrak{h}}=-\sum_p F_p.
\end{align*}

\subsubsection{Twists of surface code}
As depicted in Fig. \ref{fig:boundary} (f), there is a dislocation in the lattice, along a line $W$ (referred to as a one dimensional domain wall). Plaquettes are shifted such that the plaquette in the vicinity of $W$ is changed. $W$ can be regarded as a mixed one dimensional defect, and the point between smooth and rough  1 $d$ defects is also a special kind of defect named as twist defect \cite{Bombin2010,Kitaev2012a}. Twist defect is a zero dimensional defect, which has many interesting properties.

The plaquette operators near the domain wall will change, for example $W_p=\sigma_z^5 \sigma_z^6 \sigma_z^7 \sigma_x^4$ as depicted in  Fig. \ref{fig:boundary} (f). Besides, we must introduce a new stabilizer operator $Q=\sigma_{x}^{5}\sigma_{y}^{1}\sigma_{z}^{2}\sigma_{z}^{3}\sigma_{z}^{4}$ as depicted in Fig. \ref{fig:boundary} (f). It is easy to check that each $W_p$ and $Q$ are commutative with bulk vertex operators and plaquette operators. Therefore, we have the following twist Hamiltonian
\begin{equation}\label{eq:twistH}
H_{twist}=-\sum_{p}W_p-Q.
\end{equation}

We see that geometric change of lattice implies significant change of the Hamiltonian. If we move one $e$ particle around the point $Q$, it becomes $m$ particle, similarly for $m$ particle around $Q$. $m$ particle will condense to vacuum as moving into the smooth part of $W$, $e$ particle will condense as moving into the rough part $W$, but both $e$ and $m$ particle will condense into vacuum as moving into twist point $Q$.

\section{Neural network ansatz}
\label{sec:RBMreview}
The restricted Boltzmann machine (RBM), a shallow generative stochastic artificial neural network that can learn a probability distribution over its set of inputs, was initially invented by Smolensky~\cite{smolensky1986information} in 1986. It is a particular kind of Boltzmann machine~\cite{hinton1983optimal,ackley1985learning}. It is recently introduced in many-body physics to efficiently represent the ground state of gapped many-body quantum system \cite{Carleo602}. The approach based on RBM, the counterpart of deep neural network representation is also developed later \cite{gao2017efficient}.

We now briefly introduce the machine learning representation of a state based on restricted Boltzmann architecture.  Consider an $n$-spin physical system $\mathcal{S}=\{\mathcal{S}_1,\cdots,\mathcal{S}_n\}$, a RBM neural network contains two layers: visible layer and hidden layer (see Fig. \ref{fig:RBM}), we place $n$ spin variables $\{v_1, \cdots, v_n\}$ in a fixed basis $\{|\mathbf{v}\rangle=|v_1,\cdots,v_n\rangle\}$ on $n$ corresponding neurons in the visible layer, and there are $m$ auxiliary variables $\{h_1,\cdots,h_m\}$ where $v_i,h_j\in{\pm 1}$ in hidden layer. The neurons in visible layer are connected with the neurons in hidden layer, but there is no intralayer connections. The weights for visible neuron $v_i$  and hidden neuron $h_j$ are denoted as $a_i$ and $b_j$ respectively,  and the weight on edge between $h_j$  and $v_i$ is denoted as $W_{ji}$. Note that $\Omega=\{\mathbf{a}=(a_1,\cdots,a_n),\mathbf{b}=(b_1,\cdots,b_m), W=(W_{ij})\}$ are the parameters need to be trained  which completely determine the corresponding RBM construction. A RBM state (up to some normalization constant) is then of the form
\begin{equation}\label{eq:RBM-state}
|\Psi\rangle_{RBM}=\sum_{\mathbf{v}}\Psi(\mathbf{v},\Omega)|\mathbf{v}\rangle,
\end{equation}
where $\{\mathbf{v}\}$ is the chosen basis and the coefficient $\Psi(\mathbf{v},\Omega)$ is obtained by tracing out the hidden neuron variables \cite{Carleo602}:
\begin{align}\label{eq:RBM-coefficient}
  \Psi(\mathbf{v},\Omega)&=\sum_{\mathbf{h}}e^{\mathbf{a}^T\mathbf{v}+\mathbf{b}^T\mathbf{h}+\mathbf{h}^T W\mathbf{v}}\nonumber\\
  &= \sum_{\mathbf{h}}e^{\sum_{i}a_iv_i+\sum_{j}b_jh_j+\sum_{i,j}h_jW_{ji}v_i},\nonumber\\
   &= e^{\sum_{i}a_iv_i}\prod_{j=1}^{m}2\mathrm{cosh}(b_j+\sum_{i}W_{ji}v_i).
\end{align}

Hereinafter, we will choose $\sigma_z$ basis for each spin space, and $|+1\rangle$ and $|-1\rangle$ are two basis states such that $\sigma_{z}|+1\rangle=+1|+1\rangle$ and $\sigma_{z}|-1\rangle=-1|-1\rangle$, i.e., $\sigma_z^i|v_i\rangle =v_i|v_i\rangle$. Similarly, we have $\sigma_x^i|v_i\rangle =|-v_i\rangle$ and $\sigma_y^i|v_i\rangle =iv_i|-v_i\rangle$.

One of the most central problems of the RBM representation of quantum many-body states is its representational power. The mathematical foundation of the neural network representations is originated from representation theorem developed by Kolmogorov \cite{kolmogorov1956representation,kolmogorov1957representation} and Arnold \cite{arnold2009vladimir}, and Le Roux and Bengio's work \cite{Le2008} which stress the case for RBM.
It has been shown RBM can efficiently represent toric code states \cite{Deng2017a}, $1d$ symmetry protected topological states \cite{Deng2017a}, graph states \cite{gao2017efficient}. The connection between neural network states and tensor network states is also extensively explored \cite{gao2017efficient,huang2017neural,Chen2018,Glasser2018}. See Ref. \cite{jia2018quantum} for a review of the quantum neural network states.

\section{Neural network representation of states in stabilizer formalism}
\label{sec:ML-rep}
It is believed that RBM can represent the ground state of local gapped system. Here, we analyze RBM representations in stabilizer formalism in a much more general way. As we will see, since there is no intralayer connection in RBM, the concept of locality does not emerge. Even for some nonlocal stabilizer group, the corresponding ground state can be efficiently represented using RBM.

To begin with, we introduce our general methodology of constructing RBM representations of stabilizer states. Suppose that the stabilizer group is generated by $\{T_1,\cdots,T_m\}$. Since all other operators are just products of the generator operators, to give the stabilizer state, we only need to restrict
\begin{equation}\label{eq:stabilizer1}
T_k|\Psi\rangle=+1|\Psi\rangle.
\end{equation}

$T_k$ is the product of Pauli operators, thus we suppose that
\begin{equation}\label{eq:stabilizer2}
T_k|\mathbf{v}_k,\mathbf{\tilde{v}}\rangle=\lambda_k|\mathbf{v}_k',\mathbf{\tilde{v}}\rangle
\end{equation}
where $\mathbf{v}_k$ are the spins that $T_k$ acts nontrivially on, $\mathbf{\tilde{v}}$ are the rest of the spins, and $\lambda_k$ is the possible phase shift caused by $T_k$.

Using Eqs. (\ref{eq:stabilizer1}) and (\ref{eq:stabilizer2}) and plugging in $|\Psi\rangle=\sum_{\mathbf{v}}\Psi(\mathbf{v};\Omega)|\mathbf{v}\rangle$, we have:

\begin{align}\label{eq:general}
T_k|\Psi\rangle&=\sum_{\mathbf{v}}\Psi(\mathbf{v}_k,\mathbf{\tilde{v}};\Omega)T_k|\mathbf{v}_k,\mathbf{\tilde{v}}\rangle\nonumber\\
&=\sum_{\mathbf{v}}\Psi(\mathbf{v}_k,\mathbf{\tilde{v}};\Omega)\lambda_k|\mathbf{v}_k',\mathbf{\tilde{v}}\rangle
\end{align}

Meanwhile,
\begin{equation}\label{eq:general2}
|\Psi\rangle=\sum_{\mathbf{v}}\Psi(\mathbf{v}_k',\mathbf{\tilde{v}};\Omega)|\mathbf{v}_k',\mathbf{\tilde{v}}\rangle
\end{equation}

From Eqs. (\ref{eq:general}) and (\ref{eq:general2}) we conclude that
\begin{equation}\label{eq:generalres}
\lambda_k\Psi(\mathbf{v}_k,\mathbf{\tilde{v}};\Omega)=\Psi(\mathbf{v}_k',\mathbf{\tilde{v}};\Omega)
\end{equation}

Eq. (\ref{eq:generalres}) must hold for all spin configurations, and is almost impossible to solve directly for large systems. To tackle this problem, we employ the Jastrow wave function \cite{Jastrow1955} as a trial function and restrict that the wave function takes the form:

\begin{equation}
\Psi(\mathbf{v};\Omega)=\prod_k f_k(\mathbf{v}_k)\label{eq:prod}
\end{equation}
where each $f_k(\mathbf{v}_k)$ is the function of several local spins in the big system. We will first work out the values of $f_k(\mathbf{v}_k)$, then use hidden neuron connections to represent them.

Plugging Eq. (\ref{eq:prod}) into Eq. (\ref{eq:generalres}), we have
\begin{align*}
\lambda_k\Psi(\mathbf{v}_k,\mathbf{\tilde{v}};\Omega)&=\lambda_k\prod_k f_k(\mathbf{v}_k)\nonumber\\
&=\lambda_k f_k(\mathbf{v}_k)\prod_{l\neq k} f_l(\mathbf{v}_l)
\end{align*}
and
\begin{equation*}
\Psi(\mathbf{v}_k',\mathbf{\tilde{v}};\Omega)=f_k(\mathbf{v}_k')\prod_{l\neq k} f_l(\mathbf{v}_l').
\end{equation*}

Thus,
\begin{equation}
\lambda_k f_k(\mathbf{v}_k)\prod_{l\neq k} f_l(\mathbf{v}_l)=f_k(\mathbf{v}_k')\prod_{l\neq k} f_l(\mathbf{v}_l')\label{eq:fk}
\end{equation}

Finally, to find a solution to Eq. (\ref{eq:fk}), we let the corresponding terms equal to each other:

\begin{equation}
\lambda_k f_k(\mathbf{v}_k)=f_k(\mathbf{v}_k')\label{eq:fk1}
\end{equation}
\begin{equation}
f_l(\mathbf{v}_l)=f_l(\mathbf{v}_l')\label{eq:fk2}
\end{equation}

Given a set of stabilizers, we can find a set of $f_k(\mathbf{v}_k)$ that satisfies Eqs. (\ref{eq:fk1}) and (\ref{eq:fk2}), and then find the RBM parameters corresponding to each $f_k(\mathbf{v}_k)$.

Here, we argue that the most important thing is the stabilizer's configuration which determines the architecture of the neural network. To begin with, we divide group of stabilizer generators into several types: $\mathbf{S}_{X}$, $\mathbf{S}_{Y}$, $\mathbf{S}_{Z}$, which only contain tensor products of $\sigma_x$, $\sigma_y$ and $\sigma_z$ respectively; $\mathbf{S}_{XY}$, $\mathbf{S}_{YZ}$, $\mathbf{S}_{XZ}$, which contain tensor products of $\sigma_x$ and $\sigma_y$, of $\sigma_y$ and $\sigma_z$ and of $\sigma_x$ and $\sigma_z$; and $\mathbf{S}_{XYZ}$ which only contains tensor products of $\sigma_x$, $\sigma_y$, and $\sigma_z$. We will use the notation $\mathbf{S}_X \sqcup \mathbf{S}_Z$ to mean that the generators of the stabilizer group only involves elements of $\mathbf{S}_X$ and $\mathbf{S}_Z$ type, and similar for others.

We will prove that all code states in $\mathbf{S}_X$ (resp. $\mathbf{S}_Y$, $\mathbf{S}_Z$ ) stabilizer formalism can be exactly and efficiently represented by RBM. Specifically, we can assign one hidden neuron to each stabilizer operator which only connects with visible  neurons it acts nontrivially on, (corresponding to one $f_k(\mathbf{v}_k)$ in Eq. (\ref{eq:prod})). As for code states in $\mathbf{S}_{XZ}$ (resp. $\mathbf{S}_{XY}$, $\mathbf{S}_{YZ}$ ) stabilizer formalism, using machine learning techniques, we can give efficient RBM representation with high accuracy.

\subsection{$\mathbf{S}_X$, $\mathbf{S}_Y$ and $\mathbf{S}_Z$}
\label{sec:xyz}

Eqs. (\ref{eq:fk1}) and (\ref{eq:fk2}) suggest that we should treat each $f_k(\mathbf{v}_k)$ (i.e., each stabilizer operator) individually.

To begin with, we draw the spins $\mathbf{v}_k$ from the whole system and analyse this subsystem. In general, there will be multiple stabilizers acting on the subsystem $\mathbf{v}_k$. We will call $T_k$ the "major stabilizer" in the subsystem $\mathbf{v}_k$. The equation $T_k|\Psi\rangle=+1|\Psi\rangle$ simply corresponds to Eq. (\ref{eq:fk1}), but for other stabilizers $T_l$, the equation $T_l|\Psi\rangle=+1|\Psi\rangle$ does not correspond to Eq. (\ref{eq:fk2}). Namely, the effect of $T_l$ on $|\Psi\rangle$ is split into two parts: the possible phase shift $\lambda_l$, which is only shown in Eq. (\ref{eq:fk1}), and the possible spin flip changing $\mathbf{v}_l$ into $\mathbf{v}_l'$, which is shown in both Eq. (\ref{eq:fk1}) and Eq. (\ref{eq:fk2}). Thus, when analyzing the subsystem $\mathbf{v}_k$, the non-major stabilizers can only flip spins and cannot affect the phase.

Back to our analysis on different types of stabilizers, as a non-major stabilizer in the subsystem, $T^z\in \mathbf{S}_Z$ has no effect on the subsystem, while $T^x\in \mathbf{S}_X$ and $T^y\in \mathbf{S}_Y$ only have the effect of flipping spins. When analysing the subsystem $\mathbf{v}_k$, we can ignore all non-major $T^z$, while regard all non-major $T^y$ as $T^x$.

In conclusion, when concerning the subsystem $\mathbf{v}_k$ only, there is one major stabilizer $T_k$ and multiple non-major stabilizers $T_{l_{\mathbf{v}_k}}\in\mathbf{S}_X$ acting on them. $f_k(\mathbf{v}_k)$ describes the common eigenstate of $\{T_k, T_{l_{\mathbf{v}_k}}\}$. Since the size of the subsystem is small in general, $f_k(\mathbf{v}_k)$ can be easily found by solving Eq. (\ref{eq:stabilizer}). Treating every stabilizer $T_k$ in the same way, we can get a set of functions $\{f_k(\mathbf{v}_k)\}$, and the wave function is given by Eq. (\ref{eq:prod}).

\begin{figure}
\includegraphics[scale=0.4]{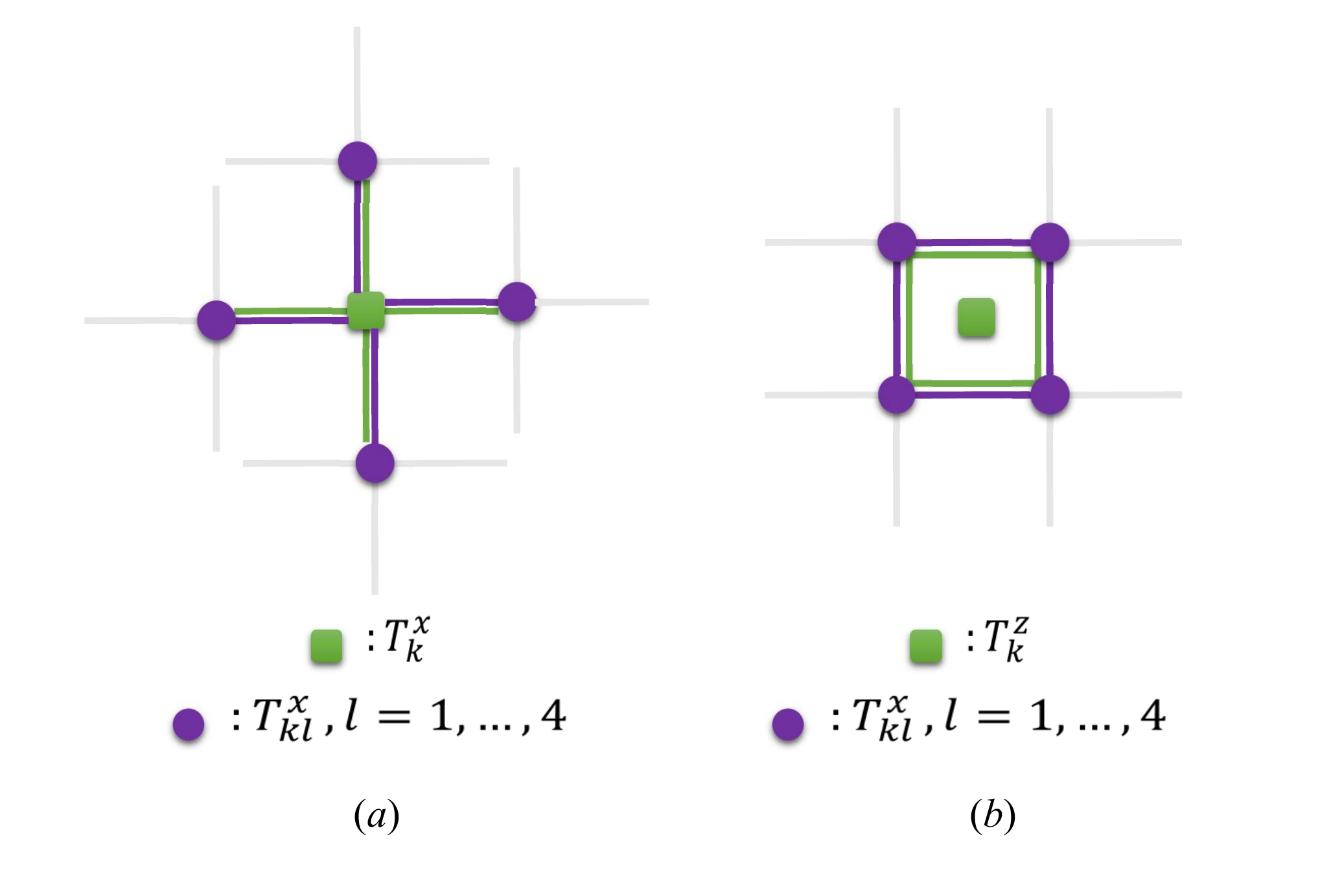}
\caption{\label{fig:vertexplaquette}$(a)$ A vertex and $(b)$ a plaquette taken from the lattice of the toric code model. Each edge corresponds to one spin, and different circles on the vertices and faces correspond to different stabilizer operators. The highlighted edges show the spins that the stabilizer acts on.}
\end{figure}

We take the Kitaev toric code state as an example, and give a much simpler and more intuitive construction compared to \cite{Deng2017a}.

Fig. \ref{fig:vertexplaquette} $(a)$ shows a vertex taken from the lattice. When concerning the four spins connected to the vertex only, there are five stabilizer operators acting on them (we ignored the four $T_z$ because of the reason stated above), with four of them independent with each other(with the relationship $T_{k1}^xT_{k2}^xT_{k3}^xT_{k4}^x=T_k^x$). We can check that $T_k^x\Psi(\mathbf{v}_k,\mathbf{\tilde{v}};\Omega)=\Psi(-\mathbf{v}_k,\mathbf{\tilde{v}};\Omega)$ is corresponding to Eq. (\ref{eq:fk1}), and the equations for $T_{k1}^x,\cdots,T_{k4}^x$ are corresponding to Eq. (\ref{eq:fk2}). The dimension $\mathcal{L}$ for this subsystem is $2^{4-4}=1$, and it is easy to find that the stabilizer state is $|++++\rangle$. For the plaquette shown in Fig. \ref{fig:vertexplaquette} $(b)$, similar results can be obtained in the same way.

Therefore, if we can construct the RBM representation of each vertex and plaquette, we will get the RBM representation of the toric code state. With the functions $f_k(\mathbf{v}_k)$, the RBM representation is easy to find. Attaching one hidden neuron to each vertex and plaquette, we can check that one solution is
$$a_k=\frac{i\pi}{4}, b_p=-i\pi, W_{pp_k}=\frac{i\pi}{4},\ \mathrm{for\ plaquettes},$$
$$a_k=0,b_q=0,W_{qq_k}=0,\ \mathrm{for\ vertices}.$$
For the general solution and details in calculation (in a less intuitive way), see Appendix A.

This result is a special case for the general $\mathbf{S}_X \sqcup \mathbf{S}_Z$ stabilizer formalism that we will discuss here.

For $T_p^z\in\mathbf{S}_Z$ and $T_q^x\in\mathbf{S}_X$, the commutation relation between them tells us that they can only share even number of spins. To begin with, we take out the spins that $T_p^z$ acts on and try to construct the function $f_p(\mathbf{v}_p)$. The equation $T_p^z\Psi(\mathbf{v}_p,\mathrm{\tilde{v}};\Omega)=\prod_p v_p \Psi(\mathbf{v}_p,\mathrm{\tilde{v}};\Omega)$ is corresponding to Eq. (\ref{eq:fk1}), and for the most general case, there exists a $T_q^x$ for any pair of spins in $\mathbf{v}_p$, which are corresponding to Eq. (\ref{eq:fk2}). Suppose there are $l$ spins that $T_p^z$ acts on, then the number of independent stabilizers is also $l$ (one $T_p^z$, and $l-1$ independent $T_q^x$), so the dimension $\mathcal{L}=2^{l-l}=1$. Therefore, we can find the unique stabilizer state of this subsystem, and express it using the function $f_p(\mathbf{v}_p)$.

Similarly, we can take out the $s$ spins that $T_q^x$ acts on and analyze this subsystem. Let $T_p^z$ and $T_{q'}^x$ be the stabilizers other than $T_q^x$ that act on part of the spins in this subsystem. $T_p^z$ do not flip spins, thus they have no effect in Eq. (\ref{eq:fk2}), and we can ignore them in the analysis. Since $T_q^x$ commutes with every other $T_{q'}^x$, in the most general case there can exist a $T_{q'}^x$ for every spin in this subsystem, and the number of independent stabilizers is $s$ (one for each spin). Therefore the dimension of this subsystem is also $1$, and we can also express the ground state of the subsystem using a function $f_q(\mathbf{v}_q)$.

Attaching one hidden neuron to each stabilizer, we can get the RBM representation of $\mathbf{S}_X \sqcup \mathbf{S}_Z$ stabilizer formalism similar to the Kitaev toric code model. One solution is
$$a_k=\frac{i\pi}{4}, b_p=-l\frac{i\pi}{4}, W_{pp_k}=\frac{i\pi}{4},\ \mathrm{for}\ T_p^z,$$
$$a_k=0,b_q=0,W_{qq_k}=0,\ \mathrm{for}\ T_q^x.$$

We can check that these solutions satisfy Eqs. (\ref{eq:fk1}) and (\ref{eq:fk2}). Hidden neurons with $b=0$ and $W=0$ have no contribution in the wave function, therefore we can remove the hidden neurons corresponding to $T_q^x$. The details in calculation can also be found in Appendix A.

\begin{algorithm}[htb]
\caption{Constructing RBM representation for $\mathbf{S}_{X}\sqcup\mathbf{S}_{Z},\mathbf{S}_{Y}\sqcup\mathbf{S}_{Z}$ and $\mathbf{S}_{X}\sqcup\mathbf{S}_{Y}$ stabilizer states}
\label{alg:Sxz}
\begin{algorithmic}[1]
\Require
The group of stabilizer generators, \\
 $G=\{T_1,T_2,\cdots,T_m\}$
\Ensure
The RBM parameters $\{a_i, b_j, W_{ij}\}$
\State Begin with no hidden neurons and all weights set to $0$.
\If{$G\in \mathbf{S}_{Y}\sqcup\mathbf{S}_{Z}$}
    \For{$v_i$ in $\mathbf{v}$}
        \State $a_i=a_i-\frac{i\pi}{4}$
        \State Replace all $\mathbf{S}_Y$ with $\mathbf{S}_X$.
    \EndFor
\ElsIf{$G\in \mathbf{S}_{X}\sqcup\mathbf{S}_{Y}$}
    \State Change to $\sigma_y$ basis.
\EndIf
\Comment{Do nothing when $G\in \mathbf{S}_{X}\sqcup\mathbf{S}_{Z}$}
\For{$j=1$ to $m$}
\If{$T_j\in \mathbf{S}_Z$}
    \State Add a hidden neuron $h_j$
        \For{$v_i\in \mathbf{v}_j$}
            \State $a_i=a_i+\frac{i\pi}{4}, b_j=b_j-\frac{i\pi}{4}, W_{ij}=\frac{i\pi}{4}$
        \EndFor
    \ElsIf{$T_j\in \mathbf{S}_X$}
        \State \textbf{continue}
    \EndIf
\EndFor
\end{algorithmic}
\end{algorithm}

However, this method fails if we try to generalize it to the situation where $T_r^y\in\mathbf{S}_{Y}$. Since $T_r^y$ commutes with every other $T_{r'}^y$, the number of spins shared by them can be odd. However, our general methodology tells us that for the subsystem $\mathbf{v}_r$, the major stabilizer is $T_r^y$, while the non-major stabilizers are actually $T_{r'_{\mathbf{v}r}}^x\in\mathbf{S}_X$, which may not commute with $T_r^y$. Therefore there may not exist a common eigenstate for the stabilizers in this subsystem, and this method fails.

The RBM representation of $T_r^y$ can be deduced from the $T_q^x$ case. Since $\sigma_x|v\rangle=|-v\rangle$ and $\sigma_y|v\rangle=iv|-v\rangle=\exp(\frac{i\pi}{2}v)|-v\rangle$, using function $T_r^y|\Psi\rangle=+1|\Psi\rangle$, we have
\begin{align*}
&T_r^y\Psi(\mathbf{v}_r,\mathbf{\tilde{v}};\Omega)|\mathbf{v}_r,\mathbf{\tilde{v}}\rangle\\
=&\exp(\frac{i\pi}{2}\sum_r v_r)T_r^x\Psi(\mathbf{v}_r,\mathbf{\tilde{v}};\Omega)|\mathbf{v}_r,\mathbf{\tilde{v}}\rangle\\
=&\Psi(-\mathbf{v}_r,\mathbf{\tilde{v}};\Omega)|-\mathbf{v}_r,\mathbf{\tilde{v}}\rangle
\end{align*}

or

\begin{align*}
&T_r^x\left[\exp(\frac{i\pi}{4}\sum_r v_r)\Psi(\mathbf{v}_r,\mathbf{\tilde{v}};\Omega)\right]|\mathbf{v}_r,\mathbf{\tilde{v}}\rangle\\
=&\left[\exp(\frac{i\pi}{4}\sum_r (-v_r))\Psi(-\mathbf{v}_r,\mathbf{\tilde{v}};\Omega)\right]|-\mathbf{v}_r,\mathbf{\tilde{v}}\rangle.
\end{align*}

Suppose the eigenstate for $T_r^x$ with eigenvalue $1$ is $|\Psi'\rangle$, then we have
$$\Psi(\mathbf{v}_r,\mathbf{\tilde{v}};\Omega)=\exp(-\frac{i\pi}{4}\sum_r v_r)\Psi'(\mathbf{v}_r,\mathbf{\tilde{v}};\Omega).$$

Therefore, we conclude that for $T_r^y$, the RBM parameters are
\begin{equation}
a_k=a_k'-\frac{i\pi}{4},b_r=b_r',W_{rr_k}=W_{rr_k}',\label{eq:Tyresult}
\end{equation}
where the parameters with a prime denotes for the parameters for the corresponding $T_r^x$ case. Note that the result here is for the whole spin system, while the results we get for $T_p^z$ and $T_q^x$ earlier are for the subsystems taken out from the big system.

Using Eq. (\ref{eq:Tyresult}) and our previous result for $\mathbf{S}_X \sqcup \mathbf{S}_Z$, we can directly get the RBM representation for $\mathbf{S}_Y \sqcup \mathbf{S}_Z$ cases. Furthermore, as for $\mathbf{S}_X \sqcup \mathbf{S}_Y$ cases, if we use $\sigma_y$ basis instead of $\sigma_z$ basis, the matrix form of the Pauli operators under the new basis reads:
\begin{equation*}
\sigma_x'=\left(
\begin{array}{cc}
0&-i\\
i&0
\end{array}
\right)\to \sigma_y
\end{equation*}
\begin{equation*}
\sigma_y'=\left(
\begin{array}{cc}
1&0\\
0&-1
\end{array}
\right)\to \sigma_z
\end{equation*}
\begin{equation*}
\sigma_z'=\left(
\begin{array}{cc}
0&1\\
1&0
\end{array}
\right)\to \sigma_x
\end{equation*}

In this way we converted $\mathbf{S}_X \sqcup \mathbf{S}_Y$ to $\mathbf{S}_Y \sqcup \mathbf{S}_Z$, which we have already solved.

To conclude, we have summarized our construction into Algorithm \ref{alg:Sxz}. However, one must note that this algorithm only specifies one code state in the entire code space. To convert between different code states, we utilize the RBM representation of string operators. See Sec. \ref{sec:excitation} for details.

\subsection{$\mathbf{S}_{XY}$, $\mathbf{S}_{XZ}$, $\mathbf{S}_{YZ}$ and $\mathbf{S}_{XYZ}$}
\label{sec:Sxz}

When generalizing our method in Sec. \ref{sec:xyz} to more complicated cases, problems arise. When $G\in\mathbf{S}_{X}\sqcup\mathbf{S}_{Z}$, for each stabilizer $T_j$, we can always take $\mathbf{v}_j$ as a subsystem, and the solution is guaranteed to exist; however, when $G\in\mathbf{S}_{XZ}$, there exist cases where there are more independent stabilizers than spins in each subsystem, so there does not exist a common eigenstate, and our method fails. Under such circumstances, merging the neighboring subsystems into bigger subsystems may help, but no general merging rules have been found, and in the worst case the size of a subsystem might reach the size of the whole system. Therefore, we resort to numerical methods instead.

Ref. \cite{jia2018RBM} proved the existence of efficient RBM representation for stabilizer states, confirming that a numerical method would work. Therefore, we try to construct a fully-connected RBM for each subsystem, and obtain the RBM representation of the state via Eq. (\ref{eq:prod}). By appropriately choosing subsystems, the system size we have to deal with will be much smaller.

Here we briefly introduce the method of training the RBM. Given the set of stabilizers $\{T_1, \cdots, T_m\}$, the Hamiltonian is defined as $H=-\sum_j T_j$, and the code states are ground states of the Hamiltonian. This problem is already solved in \cite{Carleo602}, in which they adopted reinforcement learning to minimize the energy. But for small subsystems, higher accuracy can be achieved by first calculating the full state vector $|\Psi_{j}\rangle$ for the subsystem, then minimizing the distance function
\begin{equation*}
d=\arccos\sqrt{\frac{\langle\Psi_j|\Psi_{\mathrm{RBM}}\rangle\langle\Psi_{\mathrm{RBM}}|\Psi_j\rangle}
{\langle\Psi_j|\Psi_j\rangle\langle\Psi_{\mathrm{RBM}}|\Psi_{\mathrm{RBM}}\rangle}}.
\end{equation*}

The cost for exactly computing $|\Psi_{j}\rangle$ and $|\Psi_{\mathrm{RBM}}\rangle$ grows exponentially, but is tractable for subsystems with moderate size. A small trick to reduce complexity by a constant factor is to evaluate the RBM coefficients in the sequence of the gray code.

 So far, we have given the relatively complete construction of RBM state in stabilizer formalism \cite{DuanGao}. Later in this paper we will give examples for different situations.

\begin{figure}
\includegraphics[scale=0.4]{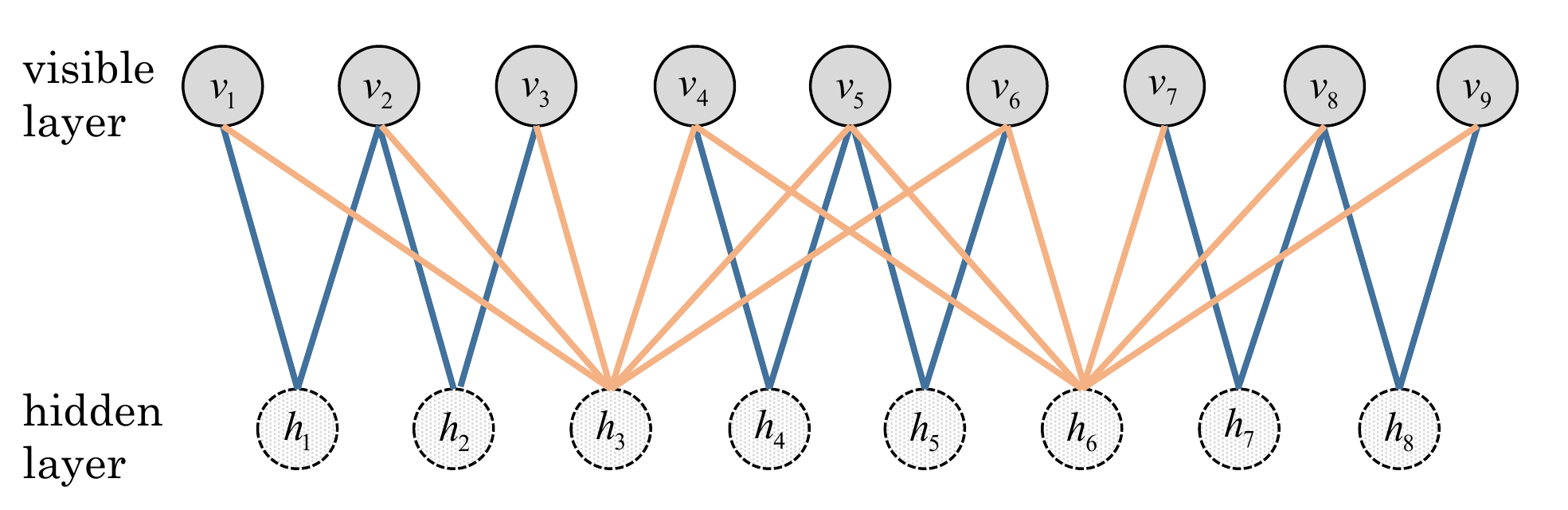}
\caption{\label{fig:RBM}Restricted Boltzmann machine representation of Shor's $[[9,1,3]]$ code state in stabilizer formalism.}
\end{figure}

Let us take Shor's $[[9,1,3]]$ code \cite{shor1995}  as an example to illustrate our general approach for constructing stabilizer states. The stabilizer generators $\mathbf{S}_{[[9,1,3]]}$ for Shor's code are
\begin{equation}\label{eq:shor-code}
\begin{split}
T_1&= \sigma_z^1\sigma_z^2IIIIIII, \\
T_2&= I\sigma_z^2\sigma_z^3IIIIII,  \\
T_3&= \sigma_x^1\sigma_x^2\sigma_x^3\sigma_x^4\sigma_x^5\sigma_x^6III, \\
T_4&=  III\sigma_z^4\sigma_z^5IIII,\\
T_5&=  IIII\sigma_z^5\sigma_z^6III,\\
T_6&= III\sigma_x^4\sigma_x^5\sigma_x^6\sigma_x^7\sigma_x^8\sigma_x^9, \\
T_7&=  IIIIII\sigma_z^7\sigma_z^8I,\\
T_8&=  IIIIIII\sigma_z^8\sigma_z^9.\\
\end{split}
\end{equation}
As depicted in Fig. \ref{fig:RBM}, we assign a hidden neuron $h_k$ to each stabilizer $T_k$ such that it is only connected with the qubits
(visible neurons) which $T_k$ acts on nontrivially. Note that $\mathbf{S}_{[[9,1,3]]}=\mathbf{S}_X\sqcup\mathbf{S}_{Z}$,  $T_1,T_2,T_4,T_5,T_7,T_8$ are of $\mathbf{S}_Z$ type with each of them acting on $l=2$ qubits nontrivially. $T_3$ and $T_6$ are of $\mathbf{S}_X$ type, and among the qubits they act on nontrivially, the number of $T_p^z$ that acted on $v_1,v_3,v_4,v_6,v_7,v_9$ is $1$, and that acted on $v_2, v_5, v_8$ is $2$. Thus the RBM parameters $\Omega_{[[9,1,3]]}$ for $|\Psi_{[[9,1,3]]}\rangle$ is
\begin{equation*}\label{}
\left\{
\begin{array}{lll}
&a_k=i\frac{\pi}{4},& k=1,3,4,6,7,9\\
&a_k=i\frac{\pi}{2}.& k=2,5,8\\
&b_p=-i\frac{\pi}{2},\,\, W_{pp_k} =i \frac{\pi}{4},& p=1,2,4,5,7,8,\\
&b_q=0,\,\,W_{qq_k} =0,&q=3,6.\\
\end{array}\right.
\end{equation*}

\section{Efficient neural network representation of surface code}
\label{sec:RMB-rep}

Using the general result obtained above, now we explicitly construct the RBM representation of defected surface code.

\subsection{Planar code with boundaries}
\begin{figure}
\includegraphics[width=9cm]{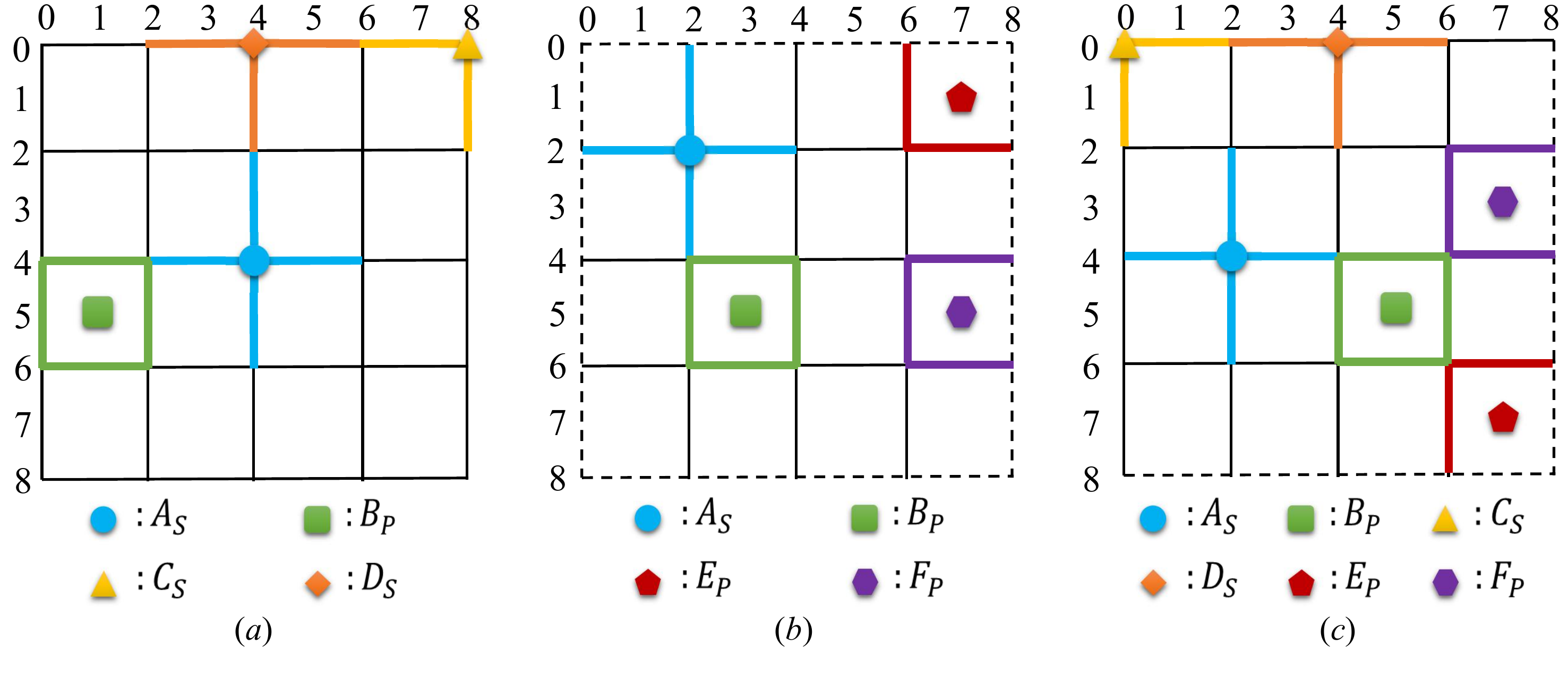}
\caption{\label{fig:boundaries}Planar code with boundaries. $(a)$ Smooth boundary; $(b)$ Rough boundary; $(c)$ Mixed boundary. The different types of stabilizers and the corresponding qubits they act on nontrivially are highlighted using different colors.}
\end{figure}

There are two types of boundaries for planar code: smooth ones and rough ones, as shown in Fig. \ref{fig:boundaries} , and we will construct RBM representation for both cases.

\subsubsection{Smooth boundaries}
We take the $4\times 4$ square lattice as a concrete example. We use $0,1,2,\cdots$ to label the rows and columns, and $X_{ij}$ for the star (plaquette) operator on vertex(face) $(i,j)$ when both $i$ and $j$ are even(odd), as shown in Fig. \ref{fig:boundaries}. Similarly, $v_{ij}$ denotes for the qubit attached to edge $(i,j)$, where $i$ and $j$ have different parity. There are four types of stabilizers:
\begin{align*}
&A_{ij}=\prod_{(m,n)\in \mathrm{star}(i,j)}\sigma_x^{mn}\\
&B_{ij}=\prod_{(m,n)\in \partial(i,j)}\sigma_z^{mn}\\
&C_{ij}=\prod_{\substack{(m,n)\in \mathrm{star}(i,j)\\(i,j)\ \mathrm{on\ the\ corner}}}\sigma_x^{mn}\\
&D_{ij}=\prod_{\substack{(m,n)\in \mathrm{star}(i,j)\\(i,j)\ \mathrm{on\ the\ boundary}}}\sigma_x^{mn}
\end{align*}
where $C_{ij}$ and $D_{ij}$ denote for the star operators on the corner and boundary, respectively. If $(i,j)$ is not on the boundary or corner, then $\mathrm{star}(i,j)$ contains the $4$ adjacent edges of vertex $(i,j)$, or  $\mathrm{star}(i,j)=\{(i-1,j),(i+1,j),(i,j-1),(i,j+1)\}$. Otherwise it contains only $3$ or $2$ adjacent edges, as depicted in Fig. \ref{fig:boundaries} $(a)$. $\partial(i,j)$ has the same expression except that $(i,j)$ denotes for a face instead of vertex. As an example, the highlighted operators in Fig. \ref{fig:boundaries} $(a)$ can be written as
\begin{equation*}
\begin{array}{ll}
A_{44}=\sigma_x^{34}\sigma_x^{54}\sigma_x^{43}\sigma_x^{45},&B_{51}=\sigma_z^{41}\sigma_z^{61}\sigma_z^{50}\sigma_z^{52},\\
C_{08}=\sigma_x^{07}\sigma_x^{18},&D_{04}=\sigma_x^{14}\sigma_x^{03}\sigma_x^{05}\\
\end{array}
\end{equation*}
Using our conclusion above, we can connect a hidden neuron to each $B_{ij}$, and the RBM parameters are:
\begin{align*}
&a_{ij}=\left\{
\begin{array}{ll}
\frac{i\pi}{2},\ &i,j\in\{1,2,\cdots,7\}\\
\frac{i\pi}{4},\ &i\in\{0,8\}\ \mathrm{or}\ j\in\{0,8\}\\
\end{array}\right.\\
&b_{B_{ij}}=-i\pi,\ W_{B_{ij},k}=\frac{i\pi}{4}\\
\end{align*}
\begin{figure}
\includegraphics[width=7cm]{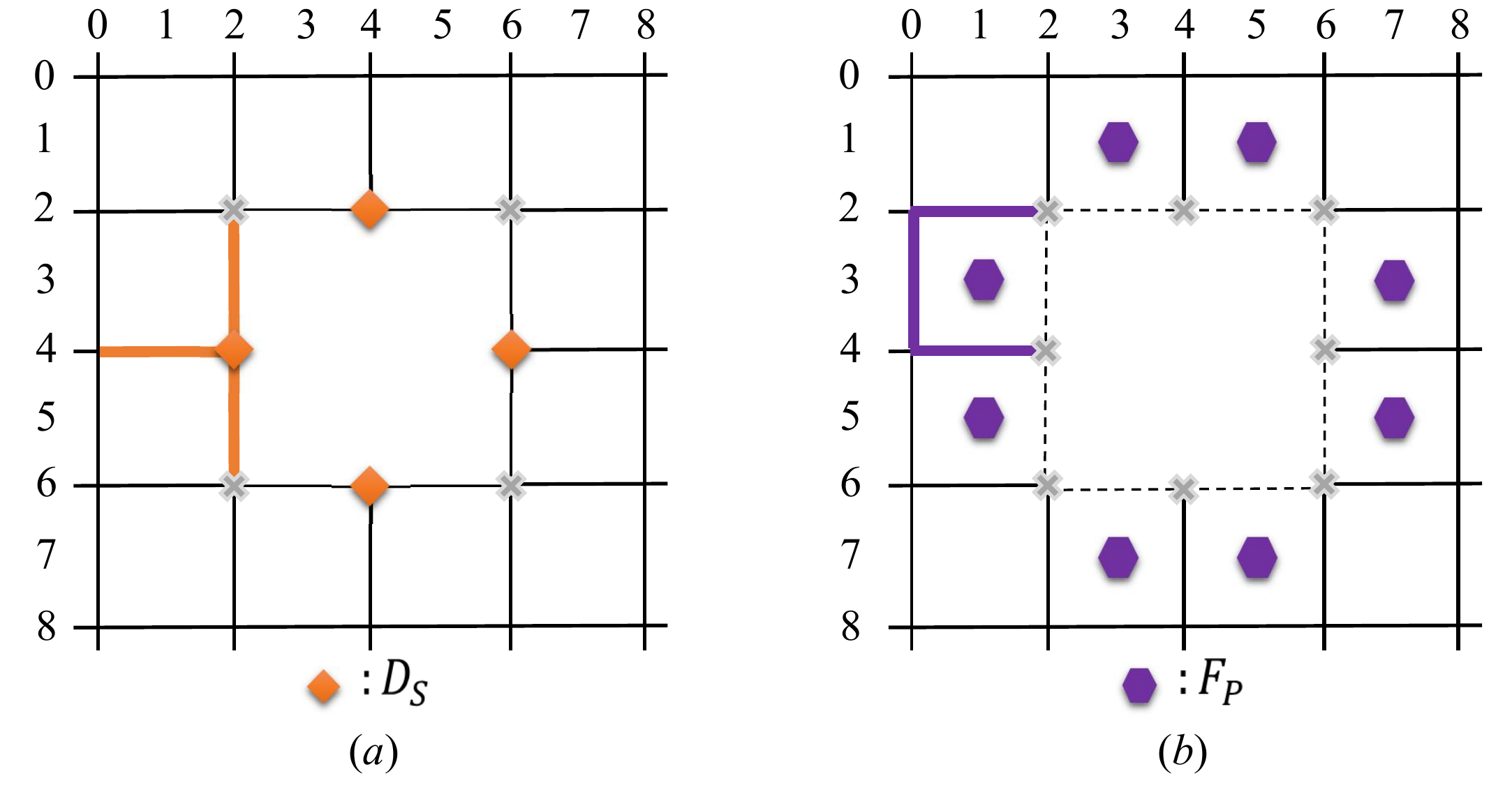}
\caption{\label{fig:defects}Planar code with defects. $(a)$ Smooth defect; $(b)$ Rough defect. The stabilizers affected by the defect are highlighted in the graph, and there are no operators defined on the vertices labeled with a cross.}
\end{figure}
\subsubsection{Rough boundaries}
We also take the $4\times 4$ square lattice as an example. There are four types of stabilizers, as shown in Fig. \ref{fig:boundaries} (b):
\begin{align*}
&A_{ij}=\prod_{(m,n)\in \mathrm{star}(i,j)}\sigma_x^{mn}\\
&B_{ij}=\prod_{(m,n)\in \partial(i,j)}\sigma_z^{mn}\\
&E_{ij}=\prod_{\substack{(m,n)\in \partial(i,j)\\(i,j)\mathrm{on\ the\ corner}}}\sigma_z^{mn}\\
&F_{ij}=\prod_{\substack{(m,n)\in \partial(i,j)\\(i,j)\mathrm{on\ the\ boundary}}}\sigma_z^{mn}
\end{align*}
As an example, the highlighted stabilizer operators are written as
\begin{equation*}
\begin{array}{ll}
A_{22}=\sigma_x^{12}\sigma_x^{32}\sigma_x^{21}\sigma_x^{23},&B_{53}=\sigma_z^{43}\sigma_z^{63}\sigma_z^{52}\sigma_z^{54},\\
E_{17}=\sigma_z^{27}\sigma_z^{16},&F_{57}=\sigma_z^{47}\sigma_z^{67}\sigma_z^{56}\\
\end{array}
\end{equation*}
Using our conclusion above, the RBM parameters for this case are:
\begin{align*}
&a_{ij}=\frac{i\pi}{2},\ b_{B_{ij}}=-i\pi,\ b_{E_{ij}}=-\frac{i\pi}{2},\ b_{F_{ij}}=-\frac{3i\pi}{4},\\
&W_{X_{ij},k}=\frac{i\pi}{4},\quad X\in\{B,E,F\}
\end{align*}

\subsubsection{Mixed boundaries}
In this example, the upper and left-hand side of the lattice have smooth boundaries, while the lower and right-hand side have rough boundaries. Therefore all six types of stabilizer appear in this example, as shown in Fig. \ref{fig:boundaries} $(c)$. We can calculate the RBM parameters in this case, which are:
\begin{align*}
&a_{ij}=\left\{
\begin{array}{ll}
\frac{i\pi}{4},\ &i=0\ \mathrm{or}\ j=0\\
\frac{i\pi}{2},\ &\mathrm{otherwise}\\
\end{array}\right.\\
&b_{B_{ij}}=-i\pi,\ b_{E_{ij}}=-\frac{i\pi}{2},\ b_{F_{ij}}=-\frac{3i\pi}{4},\\
&W_{X_{ij},k}=\frac{i\pi}{4},\quad X\in\{B,E,F\}.
\end{align*}

\subsection{Planar code with defects}
In this section, we will discuss the RBM representation of smooth and rough defects in planar code.

\subsubsection{Smooth defect}
As Fig. \ref{fig:defects} $(a)$ shows, the smooth defect causes the change in the four highlighted stabilizers, which are:
\begin{align*}
&D_{24}=\sigma_x^{14}\sigma_x^{23}\sigma_x^{25},\quad D_{42}=\sigma_x^{32}\sigma_x^{52}\sigma_x^{41},\\
&D_{46}=\sigma_x^{36}\sigma_x^{56}\sigma_x^{47},\quad D_{64}=\sigma_x^{63}\sigma_x^{65}\sigma_x^{74}.\\
\end{align*}
And the vertices $(2,2),(2,6),(6,2),(6,6)$ have no operators defined on them. Using our conclusions above, the RBM parameters in this case are:
\begin{align*}
&a_{ij}=\left\{
\begin{array}{ll}
\frac{i\pi}{4},\ &(i,j)\ \mathrm{on\ the\ boundary\ of\ the\ defect}\\
\frac{i\pi}{2},\ &\mathrm{otherwise}\\
\end{array}\right.\\
&b_{B_{ij}}=-i\pi,\ W_{B_{ij},k}=\frac{i\pi}{4}.
\end{align*}

\subsubsection{Rough defect}
As Fig. \ref{fig:defects} $(b)$ shows, the rough defect causes the change in the eight highlighted stabilizers, where $F_{31}=\sigma_z^{21}\sigma_z^{41}\sigma_z^{30}$, and similar for the others. The eight vertices labeled with a cross have no stabilizers defined on them. In this case, the RBM parameters are:
\begin{align*}
&a_{ij}=\frac{i\pi}{2},\ b_{B_{ij}}=-i\pi,\ b_{F_{ij}}=-\frac{3i\pi}{4},\\
&W_{B_{ij},k}=\frac{i\pi}{4},\ W_{F_{ij},k}=\frac{i\pi}{4}.
\end{align*}

\subsection{Planar code with twists and typical machine learning procedure for complicated cases}
\label{sec:twist}
The domain wall and twist have already been described in Sec. \ref{sec:twist}. We introduced a new twist operator $Q=\sigma_x^5\sigma_y^1\sigma_z^2\sigma_z^3\sigma_z^4$, and the plaquette operators near domain wall $W$ also changed, such as $W_p=\sigma_z^5\sigma_z^6\sigma_z^7\sigma_x^4$. Since $Q\in \mathbf{S}_{XYZ}$, and $W_p\in \mathbf{S}_{XZ}$, which we have not obtained a general result yet, in this section we explicitly construct the RBM representation of planar code with twists using machine learning techniques.
\begin{figure}
\includegraphics[scale=0.3]{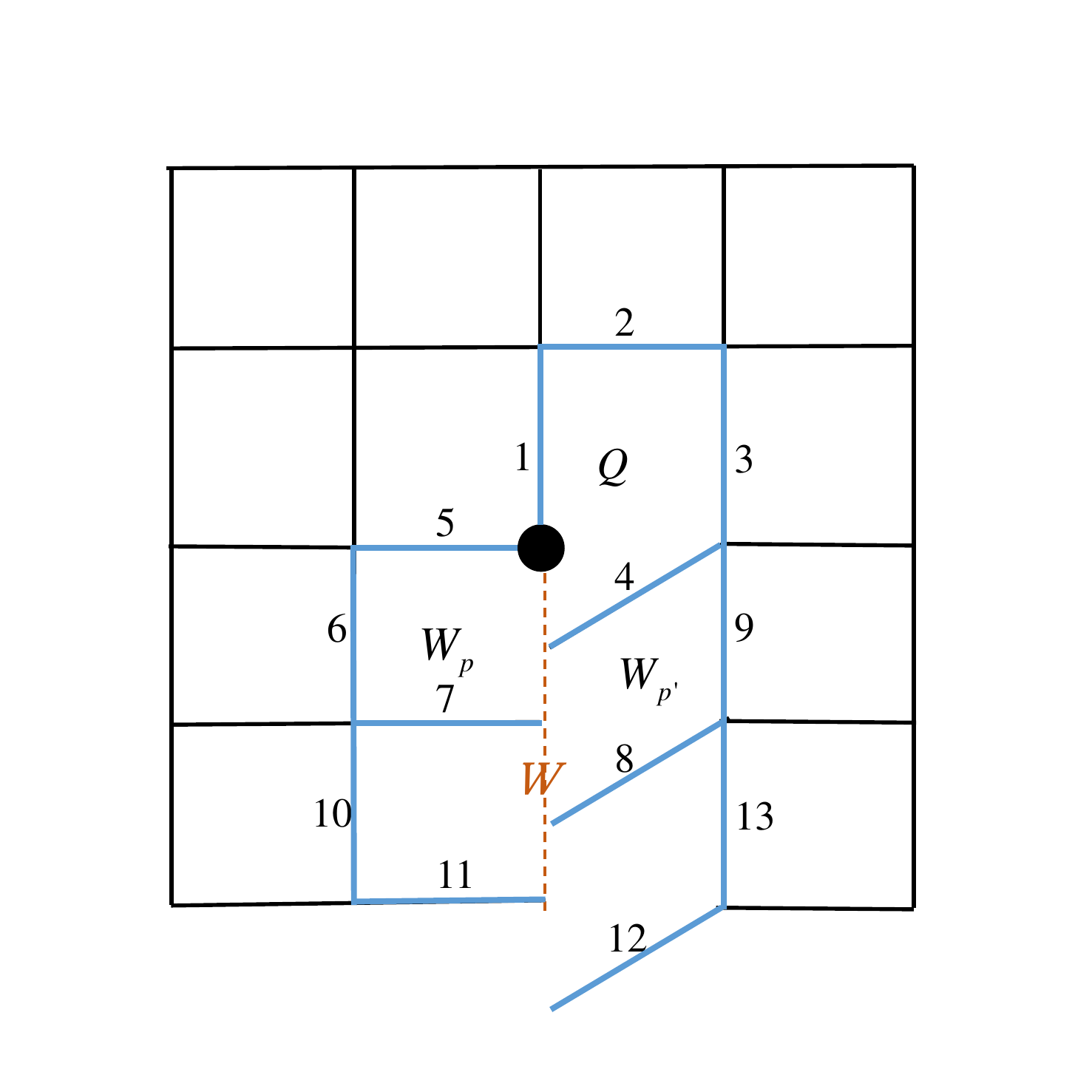}
\caption{\label{fig:twist}Planar code with a domain wall and twist.}
\end{figure}

Fig. \ref{fig:twist} shows the planar code with a domain wall and twist. As described in Sec. \ref{sec:Sxz}, we need to find a minimal subsystem in which the number of independent stabilizers is at most the same as the number of spins. It turns out that we need to include all the spins near the domain wall in the subsystem, and in this case the subsystem is the $13$ highlighted spins, with $13$ independent stabilizers acting on them. Therefore the dimension of this subsystem is $2^{13-13}=1$, so that we can find a unique ground state for it.

Then we construct a local fully connected RBM for the $13$ spins, with $13$ hidden neurons. The target state $\Phi$ is the ground state of the subsystem, and the RBM state is denoted as $\Phi'$. In the training process, we use an optimization procedure to minimize the distance function
$$d=\arccos\sqrt{\frac{\langle\Phi'|\Phi\rangle\langle\Phi|\Phi'\rangle}{\langle\Phi'|\Phi'\rangle\langle\Phi|\Phi\rangle}}.$$

\begin{figure}
\includegraphics[scale=0.27]{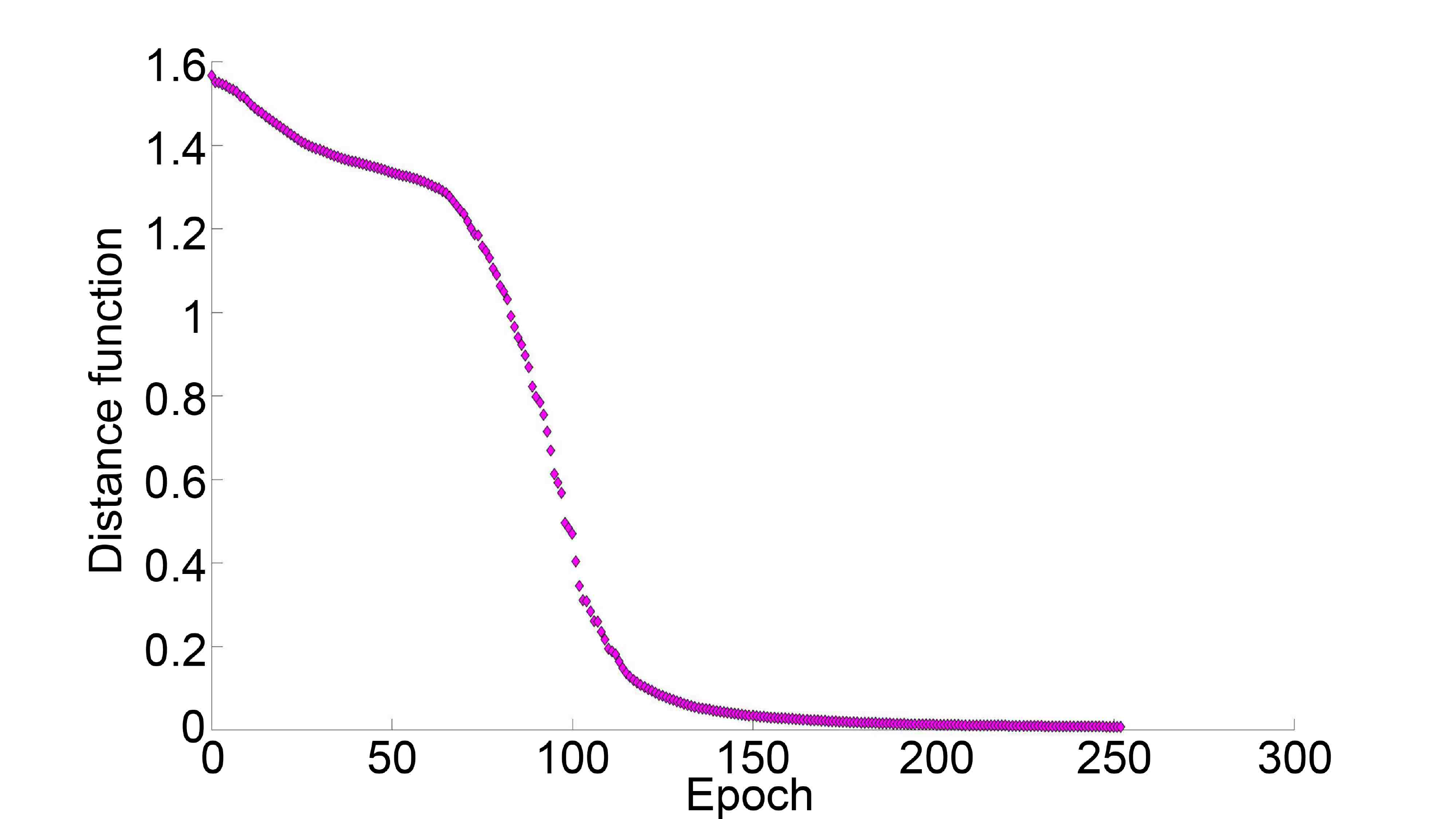}
\caption{\label{fig:twisttrain}The typical training procedure of a full connected RBM. The distance function converges smoothly to 0. }
\end{figure}

Since this system is small, we can calculate the target state $\Phi$ exactly. We used the Matlab Optimization Toolbox, which applies the Sequential Quadratic Programming (SQP) algorithm, an iterative method for nonlinear optimization, to minimize the distance function $d$ and to find a set of RBM parameters $\{a_i,b_j,W_{ij}\}$. Fig. \ref{fig:twisttrain} shows the typical optimization procedure, in which the final value of $d$ is $0.007$, indicating the fidelity is 0.99995. We can see that the distance function converges smoothly to $0$.

\subsection{Topological excitations}
\label{sec:excitation}
The RBM representation of excited states in the Kitaev toric code model has already been constructed by Deng et al. in \cite{Deng2017a}. For the completeness of our paper, we quote their results and show that edge excitation can also be represented in similar ways.

There are two types of excitations: electric excitation created by the string operator $S^z(t)=\prod_{j\in t}\sigma_z^j$, and magnetic excitation created by the string operator $S^x(t')=\prod_{j\in t'}\sigma_x^j$. Ref. \cite{Deng2017a} showed that acting the operator $S^z(t)=\prod_{j\in t}\sigma_z^j$ on the ground state is corresponding to connecting a hidden neuron $h_j$ to each $v_j$ that $S^z(t)$ acts on, with parameters $b_j=-\frac{i\pi}{2}, W_j=\frac{i\pi}{2}$. After this operation, we have
\begin{align*}
&\Psi'(\mathbf{v}_j,\mathbf{\tilde{v}})|\mathbf{v}_j,\mathbf{\tilde{v}}\rangle\\
=&\prod_j[\cosh(\frac{i\pi}{2}(v_j-1))]\Psi(\mathbf{v}_j,\mathbf{\tilde{v}})|\mathbf{v}_j,\mathbf{\tilde{v}}\rangle\\
=&\left(\prod_j\sigma_z^j\right)\Psi(\mathbf{v}_j,\mathbf{\tilde{v}})|\mathbf{v}_j,\mathbf{\tilde{v}}\rangle.
\end{align*}
And a pair of $e$ particles are created. Meanwhile, acting the operator $S^x(t')=\prod_{j\in t'}\sigma_x^j$ on the ground state is corresponding to flipping all the signs of the parameters associated to $v_j$. In this way,
\begin{align*}
&\Psi'(\mathbf{v}_j,\mathbf{\tilde{v}})|\mathbf{v}_j,\mathbf{\tilde{v}}\rangle\\
=&\Psi(-\mathbf{v}_j,\mathbf{\tilde{v}})|\mathbf{v}_j,\mathbf{\tilde{v}}\rangle\\
=&\left(\prod_j\sigma_x^j\right)\Psi(-\mathbf{v}_j,\mathbf{\tilde{v}})|-\mathbf{v}_j,\mathbf{\tilde{v}}\rangle.
\end{align*}
And a pair of $m$ particles are created.

\begin{figure}
\includegraphics[scale=0.4]{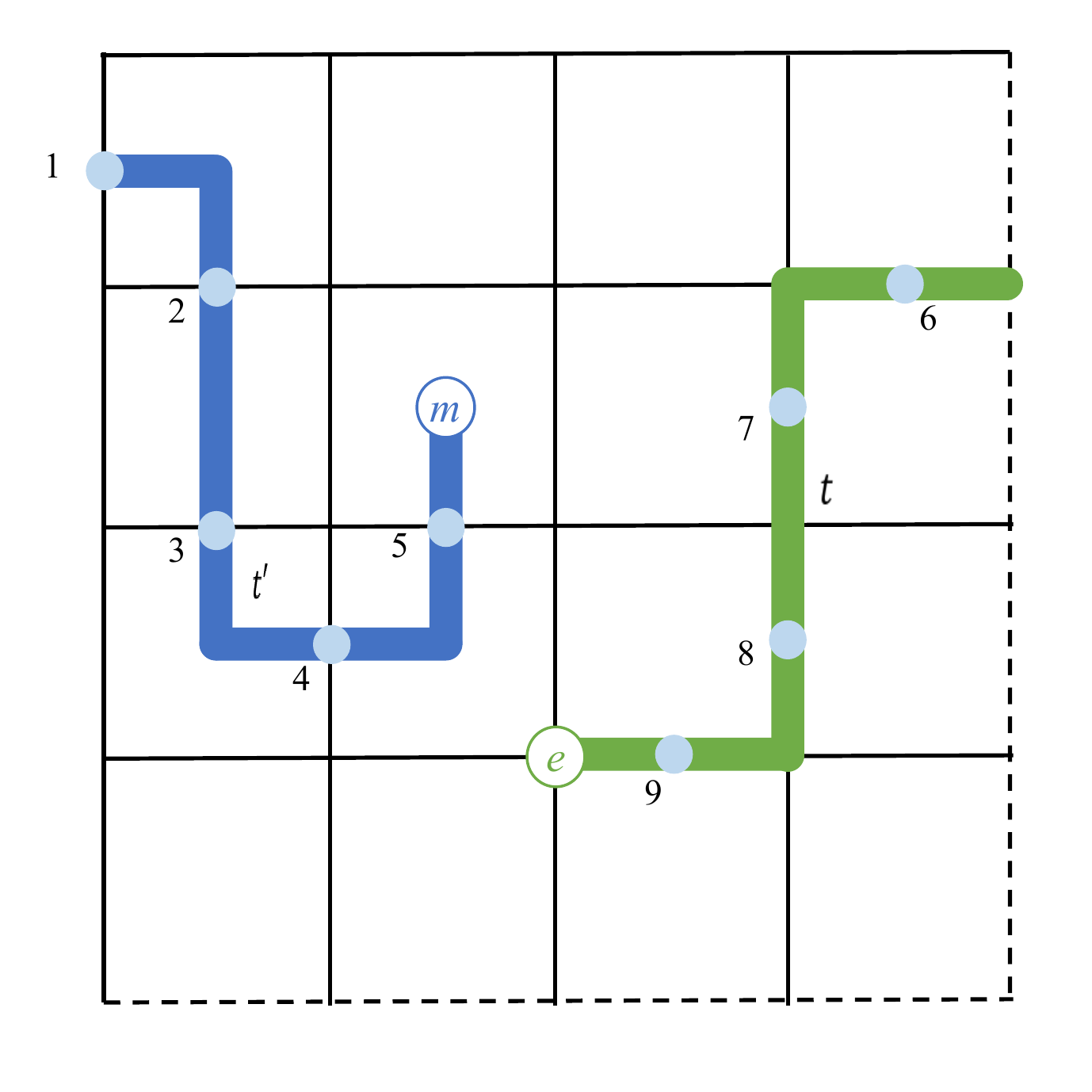}\label{fig:edge}
\caption{String operators $S^z(t)$ and $S^x(t')$. Since one end of the string operator is on the boundary, only one $e$ (or $m$) particle is created.}
\end{figure}

Fig. \ref{fig:edge} shows the two types of string operators. The string operator $S^z(t)=\sigma_z^6\sigma_z^7\sigma_z^8\sigma_z^9$ should have created a pair of $e$ particles, but since it has one end on the rough boundary, one $e$ particle condensed into vacuum as it moves into the rough boundary. Similarly, $S^x(t')=\sigma_x^1\sigma_x^2\sigma_x^3\sigma_x^4\sigma_x^5$ has one end on the smooth boundary, so that it only creates one $m$ particle on the other end. With the RBM representation of string operators $S^z(t)$ and $S^x(t')$, such physical process can be exactly and efficiently represented in RBM language.

\section{RBM representation for general $D(G)$ Kitaev model}
\label{sec:DG}
Consider a lattice with square geometry and assign $d$-level spins on each edge of the lattice. By labeling spin states with the group elements $|0\rangle,\cdots,|d-1\rangle$, we then can introduce the generalized Pauli operators
\begin{equation}\label{eq:general-Pauli}
X=\sum_{h\in\Zb_d}|h+1(\mathrm{mod}d)\rangle\langle h|,\,\,\,Z=\sum_{h\in\Zb_d}\omega^{h}|h\rangle\langle h|,
\end{equation}
where $\omega=e^{2\pi i/d}$ is the $d$-th root of unity. For the $d=2$ case, we get the usual Pauli operators $\sigma_x$ and $\sigma_z$, and they are anticommutative. In general, we have the commutation relation
\begin{equation}\label{}
ZX=\omega XZ.
\end{equation}
Since $X$ only displace the label of basis by unity, it's easy to check the eigenstates of $X$ are of the form
\begin{equation}\label{}
  |x\rangle=\frac{1}{\sqrt{d}}\sum_{h\in\Zb_d}\omega^{xh}|h\rangle,
\end{equation}
with corresponding eigenvalue $\omega^{-x}$ for each $x\in\Zb_d$.

Then we can define the star operators and plaquette operator as (see Fig. \ref{fig:surface})
\begin{equation}\label{}
  A_s =X_1X_2X^{\dagger}_3X_4^{\dagger},\,\, B_p=Z_5^{\dagger}Z_6Z_7Z_{8}^{\dagger}.
\end{equation}
Note that now the lattice is a directed graph, thus the different directions are distinguished by operators and their Hermitian conjugates. All eigenvalues of $A_v$ and $B_p$ are of the form $\omega^g$ for some $g\in \Zb_d$.

The Hamiltonian of the $D(\Zb_d)$ model is then
\begin{equation}\label{eq:ZP}
H=-\sum_{s}\sum_{h\in\Zb_d}(A_s)^h-\sum_{p}\sum_{h\in\Zb_d}(B_p)^h.
\end{equation}

Now, we try to construct the RBM representation for the general $D(G)$ Kitaev model. Since the spins can take $d$ different values, we need to generalize the traditional two-value RBM to $d$-value cases. Specifically, for the generalized RBM, the visible layer variables $\{v_1,\cdots,v_n\}$ can have $d$ different values, while the hidden layer variables $\{h_1,\cdots,h_m\}$ are still two-valued, where $v_i\in\{0,1,\cdots,d-1\}$ and $h_j\in\{+1,-1\}$. The RBM ansatz takes the same form as Eq. (\ref{eq:RBM-coefficient}), except that $v_i$ becomes $d$-valued.

To begin with, consider the equation $B_p|\Psi\rangle=+1|\Psi\rangle$. Using $|\Psi\rangle=\sum_{\mathbf{v}}\Psi(\mathbf{v};\Omega)|\mathbf{v}\rangle$, we have
\begin{align}\label{eq:DG}
&B_p\Psi(\mathbf{v};\Omega)|\mathbf{v}\rangle\nonumber\\
=&\exp(\frac{2\pi i}{d}\sum_k v_{p_k}^*)\Psi(\mathbf{v};\Omega)|\mathbf{v}\rangle\nonumber\\
=&\Psi(\mathbf{v};\Omega)|\mathbf{v}\rangle,
\end{align}
where $v_{p_k}^*=\pm v_{p_k}$, in which the plus sign is taken for the edge pointing at the positive direction (respect to the plaquette), and the minus sign for the negative direction. To make Eq. (\ref{eq:DG}) hold, we only need to restrict
\begin{equation}
\sum_k v_{p_k}^*=nd,\label{eq:DGZ}
\end{equation}
where $n$ is an integer. To this end, we connect $d-1$ hidden neurons $h_l,l\in\{1,\cdots,d-1\}$ to $\{v_{p_1},\cdots,v_{p_k}\}$, with $W_{p_l,p_k}^*=\frac{i\pi}{d}$ and $b_{p_l}=\frac{i\pi l}{d}-\frac{i\pi}{2}$. In this way, we have
\begin{align*}
\Psi(\mathbf{v};\Omega)=&\prod_p\left\{2^{d-1}\prod_l\cosh[(l+\sum_k v_{p_k}^*)\frac{i\pi}{d}-\frac{i\pi}{2}]\right\}\\
=&\prod_p\left\{2^{d-1}\prod_l\sin[(l+\sum_k v_{p_k}^*)\frac{\pi}{d}]\right\}.
\end{align*}
Since
\begin{align*}
&\prod_l\sin[(l+\sum_k v_{p_k}^*)\frac{\pi}{d}]\\
=&\left\{
\begin{array}{ll}
\pm\sin\frac{\pi}{d}\sin\frac{2\pi}{d}\cdots\sin\frac{(d-1)\pi}{d},&\sum_k v_{p_k}^*=nd\\
0,&\mathrm{otherwise}
\end{array}\right. ,
\end{align*}
we can see that this set of parameters meets our requirement.

Then let's consider the equation $A_s|\Psi\rangle=+1|\Psi\rangle$. Since $X$ is the shifting operator and each $A_s$ acts on two adjacent spins in a plaquette, we can check that if both edges point at the positive (or negative) direction (respect to the plaquette), $A_s$ will raise one spin while lowering the other; otherwise $A_s$ will raise or lower both spins. In both cases, the operator $A_s$ conserves the sum $(\sum_k v_{p_k}^* \mod d)$.

In most cases, the restriction $A_s|\Psi\rangle=+1|\Psi\rangle$ is automatically satisfied because the quantity $\sum_k v_{p_k}^*$ does not change after applying the operator $A_s$. However, $\sum_k v_{p_k}^*$ can also change by $d$, and we would have an extra $-1$ in the wave function. To make the restriction hold, we add an extra term $\exp(\frac{i\pi}{d}\sum_k v_{p_k}^*)$ to the wave function, which also adds an additional $-1$ to the wave function when $\sum_k v_{p_k}$ changes by $d$, and does not change when $\sum_k v_{p_k}^*$ does not change. We can check that $A_s|\Psi\rangle=+1|\Psi\rangle$ holds for this new wave function.

In conclusion, to represent the $D(G)$ Kitaev model in RBM language, we can connect $d-1$ hidden neurons to each plaquette, with RBM parameters
$$a_{p_k}=\pm\frac{i\pi}{d},b_{p_l}=\frac{i\pi l}{d}-\frac{i\pi}{2}, W_{p_l,p_k}=\pm\frac{i\pi}{d},$$
where for $a_{p_k}$ and $W_{p_l,p_k}$, the plus sign is taken for the edge pointing at the positive direction, and minus for the negative direction. For $d=2$, this model becomes the regular toric code model, and the RBM representation is equivalent to what we have constructed in Sec. \ref{sec:xyz} except that we use $0$ and $1$ to label spins here.

\section{Conclusions and discussions}
\label{sec:conclusion}
We have provided a systematic analysis of RBM representation in stabilizer formalism, and we find that for many crucial stabilizer groups, the exact RBM solutions exist and the number of hidden neurons is almost equal to the visible neurons. The developed results then enable us to analyze surfaces code model with boundaries, defects, domain walls and twists, and we also investigate the Kitaev's $D(\Zb_d)$ model in the form of RBM that can be optimized using variational Monte-Carlo method, with the exact solution provided. Our result sheds new light to the representational power of neural network states and gives a guidance when building the RBM neural network in stabilizer formalism. We also mention that in Ref. \cite{jia2018quantum}, we shown that all stabilizer can be reduced into the stabilizer groups which we studied in this work. Thus it is of central importance for construct RBM representation in stabilizer formalism. Many directions can been exploited further, like to provide the exact RBM solution of Kitaev's $D(G)$ model for non-Abelian group $G$ and to develop an algorithm to create RBM solution in stabilizer formalism. All these are left for our future study.

\begin{acknowledgments}
We acknowledge acknowledge Rui Zhai and Yan-Jun He for many helpful discussions. This work was supported by the National Key Research and Development Program
of China (Grant No. 2016YFA0301700), the National Natural Science Foundation of
China (Grants Nos. 11275182,11625419), and the Anhui Initiative in Quantum Information Technologies
(Grants No. AHY080000)
\end{acknowledgments}

\appendix

\section{RBM representation in stabilizer formalism}

In this appendix, we give the detailed calculation as a supplementary for Sec. \ref{sec:xyz}

For $T_{p}^{z}=\sigma_z^{p_1}\sigma_z^{p_2}\cdots\sigma_z^{p_l}\in \mathbf{S}_{Z}$, $T_{p}^{z}$ only flips the phases of spins $v_{p_1},v_{p_2},\cdots,v_{p_l}$ , i.e., $T_{p}^{z}|v_1,v_2,\cdots,v_n\rangle=(\prod_{k=1}^{l}v_{p_k})|v_1,v_2,\cdots,v_n\rangle$. Therefore the constraint $T_{p}^z|\Psi\rangle=+1|\Psi\rangle$ can be represented in RBM form as
\begin{align*}\label{}
&T_{p}^z \Psi(\mathbf{v};\Omega)|\mathbf{v}\rangle\\
=&(\prod_{k=1}^{l}v_{p_k})e^{\sum_{i}a_iv_i}\prod_{j=1}^{m}2\mathrm{cosh}(b_j+\sum_{i}W_{ji}v_i)|\mathbf{v}\rangle\\
=&e^{\sum_{i}a_iv_i}\prod_{j=1}^{m}2\mathrm{cosh}(b_j+\sum_{i}W_{ji}v_i)|\mathbf{v}\rangle.
\end{align*}
By cancelling the terms which are unrelated with the hidden neuron corresponding to $T_p$, we will get that
\begin{equation*}
(\prod_{k=1}^{l}v_{p_k})\mathrm{cosh}(b_p+\sum_{k}W_{pp_k}v_{p_k})
=\mathrm{cosh}(b_p+\sum_{k}W_{pp_k}v_{p_k})
\end{equation*}
where we use $p_k$ to label the $l$ visible neurons which are connected with $h_p$. Now if the number of $-1$ among $v_{p_k}$ is $0,2,4,\cdots$, then we further have $\mathrm{cosh}(b_p+\sum_{k}W_{pp_k}v_{p_k})=\mathrm{cosh}(b_p+\sum_{k}W_{pp_k}v_{p_k})$ which is obviously true; if the number of $-1$ among $v_{p_k}$ is $1,3,5,\cdots$, then we have $-\mathrm{cosh}(b_p+\sum_{k}W_{pp_k}v_{p_k})=\mathrm{cosh}(b_p+\sum_{k}W_{pp_k}v_{p_k})$, from which we know that $\mathrm{cosh}(b_p+\sum_{k}W_{pp_k}v_{p_k})$ must be zero. To this end, we restrict
\begin{equation}\label{eq:Z}
b_p+\sum_{k}W_{pp_k}v_{p_k}=i\frac{2m+1}{2}\pi
\end{equation}
when the number of $-1$ among $v_{p_k}$ is odd, with $m$ an integer.

There are many solutions of Eq. (\ref{eq:Z}), we need to adjust the $b_p$ and $W_{pp_k}$ to fit our need. Here we provide a solution, where we take $W_{pp_k}$ the same for all $v_{p_k}$. It is easy to check that the weights related to hidden neuron $h_p$ (which is corresponding to $T_p$) can be
\begin{equation}\label{eq:Zresult}
\left\{
\begin{array}{ll}
&b_p=-i\frac{\pi}{4},\,\, W_{pp_k} =i  \frac{\pi}{4};\,\, l=1,5,9,13\cdots\\
&b_p=-i\frac{\pi}{2},\,\, W_{pp_k} =i \frac{\pi}{4};\,\, l=2,6,10,14\cdots\\
&b_p=i\frac{\pi}{4},\,\, W_{pp_k} =i \frac{\pi}{4};\,\, l=3,7,11,15\cdots\\
&b_p=i\frac{\pi}{2},\,\, W_{pp_k} =i \frac{\pi}{4};\,\, l=4,8,12,16\cdots\\
\end{array}\right.
\end{equation}
From Eq. (\ref{eq:Z}), we can see that adding $ni\pi$ to $b_p$ will not change the result, where $n$ can be an arbitrary integer. As we only need one solution, we can rewrite Eq. (\ref{eq:Zresult}) in a more compact form:
\begin{equation}\label{eq:Zresultcompact}
b_p=-i\frac{l\pi}{4},W_{pp_k}=i\frac{\pi}{4}
\end{equation}
However, we must point out that this result only holds when there only exists one type of stabilizer $T_p^z$. More general cases will be discussed later.

The case for $T_{q}^{x}=\sigma_x^{q_1}\sigma_x^{q_2}\cdots\sigma_x^{q_s}\in \mathbf{S}_{X}$ is more complicated. $T_{q}^{x}$ will flips the spins of $v_{q_1},v_{q_2},\cdots,v_{q_l}$, i.e., $T_{q}^{x}|v_{q_1},\cdots,v_{q_s},\cdots\rangle=|-v_{q_1},\cdots,-v_{q_s},\cdots\rangle$. Therefore the constraint $T_{q}^x|\Psi\rangle=+1|\Psi\rangle$ can be represented in RBM form as
\begin{align*}\label{}
&T_{q}^{x}\Psi(\mathbf{v}_q,\mathbf{\tilde{v}};\Omega),\\
=&\Psi(-\mathbf{v}_q,\mathbf{\tilde{v}};\Omega).
\end{align*}
More precisely, if we cancel the terms unrelated to visible neurons $v_{q_k},k=1,\cdots,s$, we have
\begin{equation*}\label{eq:X}
\begin{split}
&e^{\sum_k a_{q_k}(-v_{q_k})}\mathrm{cosh}[b_q+\sum_k W_{qq_k}(-v_{q_k})] \\
&\times\prod_{q',\langle q'q\rangle}\mathrm{cosh}[b_{q'}+\sum_k W_{q'q_k}(-v_{q_k})+\sum_{q'_k\neq q_k} W_{q'q'_k}v_{q'_k}]\\
 =& e^{\sum_k a_{q_k}v_{q_k}}\mathrm{cosh}(b_q+\sum_k W_{qq_k}v_{q_k})\\
 &\times\prod_{q',\langle q'q\rangle}\mathrm{cosh}(b_{q'}+\sum_k W_{q'q_k}v_{q_k}+\sum_{q'_k\neq q_k} W_{q'q'_k}v_{q'_k}),
\end{split}
\end{equation*}
where by $\langle q'q\rangle$ we mean that $T_q$ and $T_{q'}$ share some visible neurons. To solve the equation directly is very difficult, now to illustrate the validity of our architecture, we only give one special solution to this equation, where we let the corresponding terms on each side of the equation equal to each other. And the solution can be chosen as
\begin{equation*}
a_{q_k}=ni\pi,\ W_{qq_k}=0
\end{equation*}
And $b_q$ can take any value. Specifically, we choose
\begin{equation*}
a_{q_k}=0,\ b_{q}=0,\ W_{qq_k}=0
\end{equation*}
We need to explain this result here, for it seems that we only obtained a trivial solution. Since we supposed that $T_q^x\in\mathbf{S}_{X}$, which means that all stabilizer generators only flip the spins without adding a phase factor, or that all the involved spin configurations have the same coefficient in the wave function, which is exactly the same with our result above. Therefore, we can remove the hidden neuron corresponding to $T_q^x$. Again we must emphasize that this result only holds for $T_q^x\in\mathbf{S}_{X}$, without any other types of stabilizer.

Now let us consider what will happen if we combine two sets of constraints together. To begin with we consider the case where $\{T_p^z,T_q^x\}\in\mathbf{S}_{X}\sqcup\mathbf{S}_{Z}$. $T_p^z$ does not involve spin flips, and the constraint $T_{p}^z|\Psi\rangle=+1|\Psi\rangle$ still needs to be satisfied. Thus the hidden neuron corresponding to $T_p^z$ remain unchanged, with the weights
\begin{equation}\label{eq:zresxz}
b_p=-i\frac{l\pi}{4},W_{pp_k}=i\frac{\pi}{4}
\end{equation}
However, $T_q^x$ will flip spins that $T_p^z$ acts on, and the result is different.  After cancelling the terms unrelated to $v_{q_k}$, we have
\begin{equation*}\label{eq:XZ}
\begin{split}
&e^{\sum_k a_{q_k}v_{q_k}}\mathrm{cosh}(b_q+\sum_k W_{qq_k}v_{q_k})\nonumber\\
 &\times\prod_{q',\langle q'q\rangle}\mathrm{cosh}(b_{q'}+\sum_k W_{q'q_k}v_{q_k}+\sum_{q'_k\neq q_k} W_{q'q'_k}v_{q'_k})\nonumber\\
 &\times\prod_{p,\langle pq\rangle}\mathrm{cosh}(b_{p}+\sum_k W_{pq_k}v_{q_k}+\sum_{p_k\neq q_k} W_{pp_k}v_{p_k})\nonumber\\
 =&e^{\sum_k a_{q_k}(-v_{q_k})}\mathrm{cosh}[b_q+\sum_k W_{qq_k}(-v_{q_k})] \nonumber\\
&\times\prod_{q',\langle q'q\rangle}\mathrm{cosh}[b_{q'}+\sum_k W_{q'q_k}(-v_{q_k})+\sum_{q'_k\neq q_k} W_{q'q'_k}v_{q'_k}]\nonumber\\
&\times\prod_{p,\langle pq\rangle}\mathrm{cosh}[b_{p}+\sum_k W_{pq_k}(-v_{q_k})+\sum_{p_k\neq q_k} W_{pp_k}v_{p_k}]\\
\end{split}
\end{equation*}
In order to find a solution to this equation, we first analyze the last term.
\begin{align}
&\mathrm{cosh}[b_{p}+\sum_k W_{pq_k}(-v_{q_k})+\sum_{p_k\neq q_k} W_{pp_k}v_{p_k}]\nonumber\\
=&\cosh(b_p+\sum_k W_{pp_k}v_{p_k}-2\sum_k W_{pq_k}v_{q_k})\nonumber\\
=&\cosh(b_p+\sum_k W_{pp_k}v_{p_k}-\frac{i\pi}{2}\sum_k v_{q_k})\label{eq:xzoverlap}
\end{align}
where in the last equation we used the result $W_{pq_k}=\frac{i\pi}{4}$. Since $T_p^z$ and $T_q^x$ commute with each other, the number of visible neurons shared by $T_p^z$ and $T_q^x$ is even, or $\sum_{k,\langle pq\rangle}v_{q_k}=2m$. Thus, we can further simplify Eq. (\ref{eq:xzoverlap}) into:
\begin{align*}
&\cosh(b_p+\sum_k W_{pp_k}v_{p_k}-\frac{i\pi}{2}\sum_k v_{q_k})\\
=&\left\{
\begin{array}{ll}
&\cosh(b_p+\sum_k W_{pp_k}v_{p_k}),\quad\sum_{k,\langle pq\rangle} v_{q_k}=4n\\
&-\cosh(b_p+\sum_k W_{pp_k}v_{p_k}),\ \sum_{k,\langle pq\rangle} v_{q_k}=4n+2
\end{array}\right.
\end{align*}
or
\begin{align*}
&\cosh(b_p+\sum_k W_{pp_k}v_{p_k}-\frac{i\pi}{2}\sum_k v_{q_k})\\
&=e^{\frac{i\pi}{2}\sum_{k,\langle pq\rangle}v_{q_k}}\cosh(b_p+\sum_k W_{pp_k}v_{p_k})
\end{align*}
In this way, Eq. (\ref{eq:XZ}) becomes
\begin{equation*}\label{eq:XZsimplified}
\begin{split}
&e^{\sum_k a_{q_k}v_{q_k}}\mathrm{cosh}(b_q+\sum_k W_{qq_k}v_{q_k})\nonumber\\
 &\times\prod_{q',\langle q'q\rangle}\mathrm{cosh}(b_{q'}+\sum_k W_{q'q_k}v_{q_k}+\sum_{q'_k\neq q_k} W_{q'q'_k}v_{q'_k})\nonumber\\
 =&e^{\sum_k a_{q_k}(-v_{q_k})}\mathrm{cosh}[b_q+\sum_k W_{qq_k}(-v_{q_k})] \nonumber\\
&\times\prod_{q',\langle q'q\rangle}\mathrm{cosh}[b_{q'}+\sum_k W_{q'q_k}(-v_{q_k})+\sum_{q'_k\neq q_k} W_{q'q'_k}v_{q'_k}]\nonumber\\
&\times\prod_{p,\langle pq\rangle}e^{\frac{i\pi}{2}\sum_{k,\langle pq\rangle}v_{q_k}}\\
\end{split}
\end{equation*}
To find a solution, we let
\begin{align*}
&e^{\sum_k a_{q_k}v_{q_k}}=e^{\sum_k a_{q_k}(-v_{q_k})}\prod_{p,\langle pq\rangle}e^{\frac{i\pi}{2}\sum_{k,\langle pq\rangle}v_{q_k})}\\
&\mathrm{cosh}(b_q+\sum_k W_{qq_k}v_{q_k})=\mathrm{cosh}[b_q+\sum_k W_{qq_k}(-v_{q_k})]\\
&\mathrm{cosh}(b_{q'}+\sum_k W_{q'q_k}v_{q_k}+\sum_{q'_k\neq q_k} W_{q'q'_k}v_{q'_k})\\
&=\mathrm{cosh}[b_{q'}+\sum_k W_{q'q_k}(-v_{q_k})+\sum_{q'_k\neq q_k} W_{q'q'_k}v_{q'_k}]
\end{align*}
Therefore the solution is
\begin{equation}
a_{q_k}=n_{p,q_k}\frac{i\pi}{4},\ b_q=0,\ W_{qq_k}=0,\label{eq:xresxz}
\end{equation}
where $n_{p,q_k}$ denotes for the number of $T_p^z$ that acts on $v_{q_k}$. $b_q$ can take any value, so we choose it to be $0$ and remove the hidden neuron corresponding to $T_q^x$.

To better illustrate the physical meanings of these parameters, we rearrange Eqs.(\ref{eq:zresxz}) and (\ref{eq:xresxz}):

$$a_k=\frac{i\pi}{4}, b_p=-l\frac{i\pi}{4}, W_{pp_k}=\frac{i\pi}{4},\ \mathrm{for}\ T_p^z$$
$$a_k=0,b_q=0,W_{qq_k}=0,\ \mathrm{for}\ T_q^x$$

We reassigned the parameters $a_k$, and this is the result we give in Sec. \ref{sec:xyz}. In this way, the wave function becomes
\begin{align*}
\Psi(\mathbf{v};\Omega)=&\prod_p\Bigg[\exp(\frac{i\pi}{4}\sum_{k}v_{p_k})\\
&\times 2\cosh(\frac{i\pi}{4}\sum_k (v_{p_k}-1))\Bigg]\\
=&\prod_p f_p(\mathbf{v}_p)
\end{align*}
We can check that $f_p(\mathbf{v}_p)=e^{l\frac{i\pi}{4}}$ is the same for all spin configurations with $\prod_k v_{p_k}=1$. Or, the wave function remain unchanged after flipping an even number of spins, which meets our requirement. We can further check that every condition in Sec. \ref{sec:xyz} is satisfied.
\bibliographystyle{apsrev4-1-title}

\begin{thebibliography}{54}
\bibitem [{Dua()}]{DuanGao}%
  \BibitemOpen
  \href@noop {} {\emph {\bibinfo {title} {After the work had been completed, we
  were informed that X. Gao and L.-M. Duan also develop the RBM representation
  in the stabilizer formalism, but the work have not yet be
  published.}}}\BibitemShut {Stop}%
\end{thebibliography}

\end{document}